\documentclass[apj,twocolumn,numberedappendix,appendixfloats,twocolappendix]{openjournal}

\usepackage{lipsum}

\usepackage[usenames,dvipsnames]{xcolor}
\usepackage{textgreek}
\usepackage[utf8]{inputenc}
\usepackage[english]{babel}

\usepackage{hyperref}

\hypersetup{
    unicode, 
    colorlinks=true,
    linkcolor=linkcolor,
    citecolor=linkcolor,
    filecolor=linkcolor,
    urlcolor=linkcolor,
}
\usepackage{color,colortbl}
\definecolor{linkcolor}{RGB}{28, 96, 214}
\usepackage{tensind}
\tensordelimiter{?}
\DeclareGraphicsExtensions{.bmp,.png,.jpg,.pdf}
\usepackage{verbatim}
\usepackage[normalem]{ulem}
\usepackage{orcidlink}
\usepackage{soul}

\usepackage{newtxtext,newtxmath}
\usepackage{amsmath,amsfonts,amssymb,slashed,float,bm,algorithmic,bbold,wasysym}
\usepackage{array,multirow,multirow,dcolumn,booktabs}
\usepackage{physics}
\usepackage{savesym}
\savesymbol{tablenum}
\usepackage{siunitx}
\restoresymbol{SIX}{tablenum}
\usepackage{xspace}
\usepackage[capitalise]{cleveref}

\makeatletter
\usepackage{etoolbox}
\patchcmd\H@refstepcounter{\protected@edef}{\protected@xdef}{}{}
\makeatother

\DeclareUnicodeCharacter{2212}{-}

\newcommand{\deemph}[1]{{\color{black!40}#1}}

\defcitealias{niemiec2019}{N19}
\defcitealias{darragh-ford2023}{DF23}

\begin{document}

    \title{Non-spherical BUFFALOs: a weak lensing view of the Frontier Field clusters and associated systematics}

    \author{A.~Niemiec\orcidlink{0000-0003-3791-2647}$^1$}
    \author{A.~Acebron\orcidlink{0000-0003-3108-9039}$^{2,3}$}
    \author{B.~Beauchesne\orcidlink{0000-0002-0443-6018}$^{4,5}$}
    \author{M.~Jauzac$^{4,5,6,7}$}
    \author{J.~M.~Diego$^2$}
    \author{D.~Eckert$^8$}
    \author{D.~Harvey$^9$}
    \author{A.~M.~Koekemoer\orcidlink{0000-0002-6610-2048}$^{10}$}
    \author{D.~J.~Lagattuta\orcidlink{0000-0002-7633-2883}$^{11}$}
    \author{M.~Limousin$^{12}$}
    \author{G.~Mahler\orcidlink{0000-0003-3266-2001}$^{13}$}
    \author{N.~Patel$^{4,5}$}
    \author{S.~Tam\orcidlink{000-0002-6724-833X}$^{14}$}
    \author{J.~F.~V.~Allingham\orcidlink{0000-0003-2718-8640}$^{15}$}
    \author{R.~Cen$^{16, 17}$}
    \author{A.~Faisst$^{18}$}
    \author{D.~Perera\orcidlink{000-0002-4693-0700}$^{19}$}
    \author{M.~Sereno\orcidlink{0000-0003-0302-0325}$^{20,21}$}
    
    \email{niemiec@lpsc.in2p3.fr}
    
    \affiliation{$^1$Univ. Grenoble Alpes, CNRS, Grenoble INP, LPSC-IN2P3, 38000 Grenoble, France}
    \affiliation{$^{2}$Instituto de Física de Cantabria (CSIC-UC), Avda. Los Castros s/n, 39005 Santander, Spain}
    \affiliation{$^{3}$INAF -- IASF Milano, via A. Corti 12, I-20133 Milano, Italy}
    \affiliation{$^4$Centre for Extragalactic Astronomy, Durham University, South Road, Durham DH1 3LE, UK}
    \affiliation{$^5$Institute for Computational Cosmology, Durham University, South Road, Durham DH1 3LE, UK}
    \affiliation{$^6$Astrophysics Research Centre, University of KwaZulu-Natal, Westville Campus, Durban 4041, South Africa}
    \affiliation{$^7$School of Mathematics, Statistics \& Computer Science, University of KwaZulu-Natal, Westville Campus, Durban 4041, South Africa}
    \affiliation{$^8$Department of Astronomy, University of Geneva, Ch. d’Ecogia 16, CH-1290 Versoix, Switzerland}
    \affiliation{$^9$Laboratoire d’Astrophysique, EPFL, Observatoire de Sauverny, 1290 Versoix, Switzerland}
    \affiliation{$^{10}$Space Telescope Science Institute, 3700 San Martin Drive, Baltimore, MD 21218, USA}
    \affiliation{$^{11}$Centre for Astrophysics Research, Department of Physics, Astronomy and Mathematics, University of Hertfordshire, Hatfield AL10 9AB, UK}
    \affiliation{$^{12}$Aix Marseille Univ, CNRS, CNES, LAM, Marseille, France}
    \affiliation{$^{13}$STAR Institute, Quartier Agora - All\'ee du six Ao\^ut, 19c B-4000 Li\`ege, Belgium}
    \affiliation{$^{14}$Institute of Physics, National Yang Ming Chiao Tung University, No. 1001, Daxue Rd. East Dist., Hsinchu City, Taiwan}
    \affiliation{$^{15}$Department of Physics, Ben-Gurion University of the Negev, P.O. Box 653, Be'er-Sheva 84105, Israel}
    \affiliation{$^{16}$Center for Cosmology and Computational Astrophysics, Institute for Advanced Study in Physics, Zhejiang University, Hangzhou 310027, China}
    \affiliation{$^{17}$Institute for Astronomy, School of Physics, Zhejiang University, Hangzhou 310027, China}
    \affiliation{$^{18}$Caltech/IPAC, 1200 E. California Boulevard, Pasadena, CA 91125, USA;}
    \affiliation{$^{19}$School of Physics and Astronomy, University of Minnesota, Minneapolis, MN, 55455, USA}
    \affiliation{$^{20}$INAF - Osservatorio di Astrofisica e Scienza dello Spazio di Bologna, via Piero Gobetti 93/3, I-40129 Bologna, Italy}
    \affiliation{$^{21}$INFN, Sezione di Bologna, viale Berti Pichat 6/2, I-40127 Bologna, Italy}

  \begin{abstract}
   Galaxy clusters are tracers of the large scale structures of the Universe, making the time evolution of their mass function dependent on key cosmological parameters, such as the cosmic matter density or the amplitude of density fluctuations $\sigma_8$. Accurate measurements of cluster's total masses are therefore essential, yet they can be challenging, particularly for clusters with complex morphologies, as simple mass profiles are often adopted to fit the measurements.
   In this work, we focus on the Frontier Fields galaxy clusters: a sample of six extremely massive systems, that, in most cases, exhibit highly complex mass distributions.     The BUFFALO survey extended the \emph{Hubble Space Telescope} observations for the Frontier Fields galaxy clusters, providing high-resolution multi-band imaging within a few Mpc. Combining this high-quality imaging dataset with ancillary spectroscopy, we produce weak-lensing catalogues with very high source densities, about 50 sources/arcmin$^2$. This allows us to robustly estimate the individual weak-lensing cluster masses and quantify the sensitivity of these measurements on different factors, such as the cluster centring, the uncertainty on the redshift distribution or the foreground contamination and boost factor correction. This provides a data-driven analysis of the different sources of systematics that can impact such measurements.
   We find that the largest sources of systematic bias arise for the most disturbed clusters, such as the multi-modal, merging galaxy cluster Abell 2744. This analysis sets a comprehensive framework for assessing the impact of systematics on the weak-lensing estimates of cluster masses, and in particular, in the case of unrelaxed clusters. This can play a key role in forthcoming cosmological analyses based on wide-field surveys such as \emph{Euclid} and the Legacy Survey of Space and Time of the Rubin Observatory.
   The weak lensing catalogues will be made available upon acceptance of the publication at \url{https://archive.stsci.edu/hlsp/buffalo}.
   \end{abstract}


   \maketitle

\section{Introduction}

Galaxy clusters are the most massive gravitationally bound structures in the Universe, residing at the nodes of the cosmic web. As such, they are powerful cosmological probes: their abundance as a function of mass and redshift is highly sensitive to the growth of structure and the underlying cosmological parameters, including the matter density, dark energy equation of state, and amplitude of primordial fluctuations. By tracking the evolution of the cluster mass function over cosmic time, we can place stringent constraints on cosmological models.

However, leveraging clusters for precision cosmology critically depends on accurate estimates of their total mass. Several methods exist to infer cluster masses: X-ray emission, primarily due to thermal Bremsstrahlung from the intracluster medium (ICM), allows for measurements of the gas density \citep[e.g.][]{pratt2002} and temperature \citep[e.g.][]{arnaud1996}, which can be used to estimate the total mass under the assumption of hydrostatic equilibrium \citep{sarazin1988}; the Sunyaev-Zel'dovich (SZ) effect \citep{sunyaev1972}, caused by inverse Compton scattering of Cosmic Microwave Background (CMB) photons on ICM electrons, is measurable in millimetric observations and probes the integrated electron pressure; the velocity dispersion of cluster member galaxies provides a dynamical mass estimate \citep[e.g.][]{katgert2004}; and gravitational lensing, resulting from the deflection of light by massive structures, enables the direct mapping of the matter distribution, both luminous and dark, along the line of sight. Each of these methods carries its own sources of systematic uncertainty and bias, which must be characterized and mitigated to achieve accurate mass estimates \citep[see][for a review]{pratt2019}. Among them, lensing stands out as particularly robust, as it probes the total projected mass without relying on assumptions about the cluster’s dynamical state or the physics of the intracluster gas.

Gravitational lensing can be observed in two complementary regimes: strong and weak lensing. Strong lensing occurs in the inner, high-density regions of clusters that produce highly distorted images such as arcs and multiple images of background galaxies. In clusters presenting such features, it enables detailed modelling of the central mass distribution with high precision. However, strong lensing features are relatively rare and limited to a subset of clusters. In contrast, weak lensing induces subtle, coherent distortions in the shapes of background galaxies across the entire cluster field, including the outer regions. While it offers lower spatial resolution, weak lensing is universally present and thus applicable to the full cluster population, making it especially valuable for statistical studies of large samples.

The challenge arises when scaling lensing-based mass estimates to the vast number of clusters accessible in current and upcoming wide-field surveys. Missions such as Euclid \citep{laureijs2011, euclid2022, mellier2025} and the Legacy Survey of Space and Time (LSST) at the Rubin Observatory \citep{ivezic2019} are expected to detect tens to hundreds of thousands of clusters, making detailed mass modelling of individual systems impractical. In this context, cluster masses are typically inferred by fitting simple parametric models, such as Navarro-Frenk-White \citep[NFW][]{nfw1996} profiles, to azimuthally averaged weak lensing shear profiles, under simplifying assumptions such as spherical symmetry and unimodal mass distributions.

While this approach performs well for relaxed and morphologically regular clusters, its accuracy becomes less certain in the presence of complex or disturbed systems. Substructures, triaxial mass distributions, and line-of-sight projections can all cause discrepancies between the observed lensing signal and the assumed parametric model, introducing model bias in the inferred masses. In most analyses, stacking techniques are employed to boost the signal-to-noise ratio, since the weak-lensing signal of individual clusters is often too low to be measured with high significance. Stacking can help mitigate the impact of irregularities by averaging over large ensembles, but if a substantial fraction of the sample systematically deviates from the assumed model, residual biases may persist and propagate into cosmological constraints. These issues are expected to become more pronounced at higher redshifts, where clusters are more likely to be dynamically young, actively forming, and undergoing mergers.

In this context, it is crucial to assess the reliability of standard shear-based mass estimate methods across the full diversity of cluster morphologies. As a pilot study, we focus here on a sample of six extremely massive and morphologically complex clusters: Abell\,2744, Abell\,370, Abell\,S1063, MACS\,J0416.1$-$2403, MACS\,J0717.5$+$3745 and MACS\,J1149.5$+$2223 (hereafter A2774, A370, AS1063, M0416, M0717 and M1149, respectively). These clusters were originally selected for the \emph{ Hubble Frontier Fields} program \citep[HFF,][]{lotz2017} due to their exceptional lensing strength. As such, their central regions have been observed to unprecedented depth with the \emph{Hubble Space Telescope} (\emph{HST}). These deep observations were later extended by the BUFFALO program \citep[\emph{Beyond Ultra-deep Frontier Fields And Legacy Observations}][]{steinhardt2020}, which targeted the surrounding, less dense regions with the \emph{HST} to enable high-resolution weak lensing measurements across a broader field of view. The BUFFALO dataset provides a high density of background sources, allowing for the measurement of individual shear profiles around each cluster. We use these data to estimate the weak-lensing masses of the six clusters, applying standard parametric fitting methods. Given their high masses and complex dynamical states, this sample offers a unique opportunity to test the impact of simplifying assumptions, such as spherical symmetry or smoothness, on the estimated mass values, and to quantify the potential biases that may arise in similarly complex systems in upcoming wide-field surveys.

This paper is organized as follows: in Sect.\ref{sec:obs}, we describe the datasets used in this analysis. The construction of the weak-lensing catalogues is detailed in Sect.\ref{sec:wl_cats}. In Sect.\ref{sec:wl_masses}, we present the measurements of the fiducial weak-lensing-only masses, as well as the combined strong+weak lensing mass estimates. We explore the impact of different modelling assumptions on mass estimation in Sect.\ref{sec:discussion}. Throughout the paper, we adopt the Planck 2018 $\Lambda$CDM cosmology \citep{planck2018}. All distances are expressed in comoving units, $\log$ refers to the base-10 logarithm, and $\ln$ to the natural logarithm.

\section{Observations}
\label{sec:obs}

    \subsection{HST}

The BUFFALO survey \citep[GO-15117; PIs: Steinhardt and Jauzac,][]{steinhardt2020} was constructed around the HFF \citep[GO-14038; PI: Lotz,][]{lotz2017} observations, with the aim of expanding the area observed by HST around the six galaxy cluster fields with the Advanced Camera for Survey (ACS) and the Wide Field Camera 3 (WFC3). 
The HFF campaign targeted six particularly massive clusters, presented in Table~\ref{tab:clusters}, that were primarily selected for their large lensing strength while considering a low sky brightness and Galactic extinction, availability of parallel fields, and ground-based follow-up observations. 

We refer the reader to \citet{steinhardt2020} for a detailed discussion about the BUFFALO survey design and strategy and we provide hereafter a brief summary.
The BUFFALO programme observed each of the six lens cluster fields for a total of 16 orbits, in five of the seven HST filters used for the HFF observations, i.e. ACS/F606W, ACS/F814W, WFC3/F105W, WFC3/F125W and WFC3/F160W. The BUFFALO ACS and WFC3 footprints cover three and four times the area of the original HFF ones for both the main cluster and the parallel field, respectively. In summary, the exposure depths in these extended BUFFALO regions are 1538s and 2934s for ACS F606W and F814W respectively, and 1426s, 1467s, 1579s for WFC3/IR F105W, F125W, F160W respectively, in all cases obtained using a 4-point dither pattern optimized for half-pixel subsampling in both ACS and WFC3/IR for the prime and parallel exposures.

\begin{figure*}
    \begin{center}
    \hspace{-5mm}\includegraphics[width=\textwidth,angle=0.0]{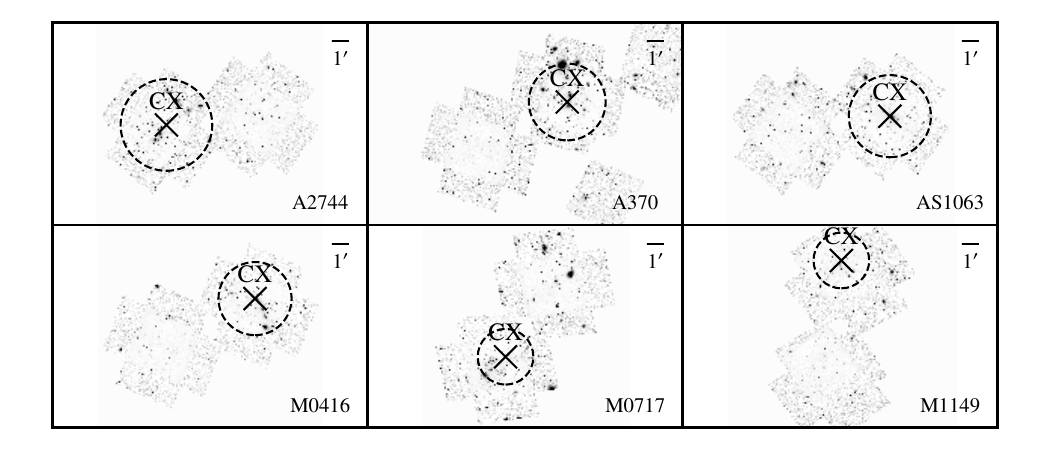}
    \caption{BUFFALO observations in the ACS/F814W band for the six galaxy clusters (which also incorporates the data taken within the HFF and other ancillary programmes). The fiducial cluster centres, corresponding to the X-ray surface-brightness peaks, are shown as black crosses, while the circles have a radius of 1 comoving Mpc. The mosaic pattern of the BUFFALO observations can be seen on the image, with the two adjacent fields for each cluster, the main and the parallel, each $\sim 5 ~\rm arcmin^2$. Note the small gap in the mosaic between the two fields.}
    \label{fig:BUFFALO}
    \end{center}
\end{figure*}

All the HST imaging data (i.e. including both the BUFFALO and the other overlapping HST data) were astrometrically aligned relative to each other and to the {\em Gaia}\footnote{\url{https://www.cosmos.esa.int/web/gaia/}} absolute astrometric frame, and combined into mosaics with a pixel size of 30\,mas and 60\,mas for ACS and WFC3 respectively, following the approaches described in \citet{koekemoer2011}, which go beyond the standard STScI pipeline processing in particular using improved detector calibrations for dark current, bias, flatfields, and other detector-level improvements, as well as improved astrometric alignment.
We also make use of the individual exposures in order to facilitate PSF characterization for lensing measurements. All the data products from the BUFFALO survey are available at the Mikulski Archive for Space Telescopes (MAST) as High Level Science Products\footnote{The data products can be accessed through \url{https://archive.stsci.edu/hlsp/buffalo}.}.
The full BUFFALO ACS/F814W mosaics of the six cluster fields are shown in Fig~\ref{fig:BUFFALO}.

 \begin{table}
    \centering
    \begin{tabular}{c|c c c}
Galaxy cluster  & RA$_{\rm{ref}}$   & Dec$_{\rm{ref}}$  & $z_{\rm{cl}}$ \\
\hline
A2744     & 3.5799            &  $-$30.3912       & 0.308 \\  
A370      & 39.9725           & $-$1.5789         & 0.375 \\  
AS1063    & 342.1852          & $-$44.5301        & 0.347 \\ 
M0416     & 64.0395           & $-$24.0670        & 0.397 \\ 
M0717     & 109.3798          & 37.7584           & 0.540 \\ 
M1149     & 177.3970          & 22.4020           & 0.544 \\ 
    \end{tabular}
    \caption{The BUFFALO/Frontier Fields clusters, with X-ray surface brightness peak positions, and redshifts.}
    \label{tab:clusters}
\end{table} 

    \subsection{Ancillary X-ray observations}

\begin{table}
    \centering
    \begin{tabular}{c|c c }
Galaxy  & Exposure& Observation IDs \\
cluster & time (ks)& \\
\hline
A2744     & 100 & 8557, 8477, 7915, 7712\\
A370      & 133 & 0782150101 \\
AS1063    & 123 & 4966, 18818, 18611 \\
M0416     & 187 & 10446, 16236, 16523, 17313\\
M0717     & 242 & 1655, 4200, 16235, 16305 \\
M1149     & 363 &1656, 3589, 16238, 16239, 17595\\
&&  17596, 16306, 16582\\
    \end{tabular}
    \caption{Summary of the X-ray observations used in this work with the combined exposure time among all observations and the associated ID. Besides A370 where we use \textit{XMM-Newton} observations, all the others are from the \textit{Chandra X-ray Observatory}.}
    \label{tab:X-ray-obs}
\end{table} 

To complement these optical observations, we use for some aspects of our analysis X-ray observations of the clusters. Table~\ref{tab:X-ray-obs} shows the combined exposure time of each considered observation per cluster with the associated ID. \textit{Chandra X-ray Observatory} data are used for most of the clusters, with the exception of A370, for which we use \textit{XMM-Newton} observations. For this cluster, the data was reduced as described in \citet{ghirardini2019, rossetti2024} using XMMSAS v19.1 and the analysis pipeline developed in the framework of the \emph{XMM-Newton} cluster outskirts project \citep[X-COP,][]{xcop}, and we use the count map produced in the [0.5-2] keV band.  In the case of data from the \textit{Chandra X-ray Observatory}, we reduce the data following the procedure detailed in \citet{beauchesne2023} with the \textit{Chandra} pipeline \textsc{ciao}\footnote{\url{https://cxc.cfa.harvard.edu/ciao/}} $4.15$ \citep{ciao2006} and \textsc{caldb} $4.10.7$. We produced count maps in the broad energy band (i.e. [0.5,7] keV) that are used in the rest of this analysis. Our procedure includes the removal of point sources from the imaging data.

In particular, we use the X-ray surface brightness maps to define a fiducial centre for each galaxy cluster. Because of the asymmetric distribution of most systems in the sample, we use as centre the surface brightness peaks, computed using the \textsc{pyproffit} package\footnote{\url{https://pyproffit.readthedocs.io/en/latest/}} \citep{eckert2020}. The corresponding positions are given in Table~\ref{tab:clusters}, and shown as black crosses in Fig.~\ref{fig:BUFFALO}. We examine in details the impact of this choice for the cluster centre in Sect.~\ref{sec:miscentring}.

\section{Weak lensing catalogues}
\label{sec:wl_cats}

In the weak lensing regime, the shape of background galaxies is only slightly distorted by the gravitational potential of the lens. The mapping between the unlensed and lensed coordinates can therefore be locally linearised, and described by the Jacobian matrix of the lens equation, called the amplification matrix:
\begin{equation}
    A = 
    \begin{pmatrix}
        1 - \kappa - \gamma_1   & -\gamma_2 \\
        -\gamma_2               & 1 - \kappa + \gamma_1
    \end{pmatrix},
\end{equation}
where the convergence, $\kappa$, describes the isotropic magnification of the lensed image, and the shear, $\gamma$, its anisotropic distortion.
The reduced shear, defined as $g = \frac{\gamma}{1 - \kappa}$, is therefore an observable of the gravitational potential of the lens, and it is encoded in the observed shape of the weakly lensed background sources. Measuring galaxy ellipticity to extract the shear is a critical step in the analysis, and presents a number of challenges, including, for instance, the precise measurement of the ellipticities themselves, the correction of instrumental and atmospheric effects (i.e the point-spread-function, or PSF, correction), the calibration of the response of the measured ellipticity to the shear, and the selection of background galaxies which are gravitationally lensed.

To obtain the weak lensing catalogues from the ACS/F814W observations for the six clusters fields and tackle these different difficulties, we apply a measurement process based on the one detailed in \citet{jauzac2012} and \citet{niemiec2023}, which can be divided into the following steps: (1) extract the sources from the coadded image, and separate stars from galaxies; (2) measure their shapes, corrected for the PSF, using both the coadded image and individual exposures; (3) clean the catalogue to remove artefacts and galaxies with ill-measured shapes; (4) select galaxies that are located behind the lens cluster and that therefore carry the weak-lensing information. Steps (1), (2) and part of step (3) were performed using the publicly available pyRRG\footnote{\url{https://pypi.org/project/pyRRG/}} code \citep{harvey2019}, which is based on a shape measurement algorithm first introduced by \cite{rhodes2000}, RRG. We briefly review these different aspects, necessary for the production of the shear catalogues.

    \subsection{pyRRG shear measurement}
    \label{sec:pyrrg}

\paragraph*{\textit{Source extraction and galaxy/star separation.}} The pyRRG pipeline first extracts all sources in the field by running the \textsc{SExtractor} software \citep{bertin1996} on the coadded ACS/F814W image, in the so-called ``hot/cold'' configuration, which optimises deblending and detection of small and faint sources. Stars and galaxies are then selected from image artefacts, using a random forest classifier. This automatic selection is then visually inspected, by projecting the sources into the (1) magnitude - $\mu_{\mathrm{max}}$ (defined as the `\textsc{mag\_auto}' and `\textsc{mu\_max}' \textsc{Source Extractor} parameters respectively); (2) magnitude - size (where the size is computed from the weighted second order moments as $(M_{xx} + M_{yy})/2$); and (3) magnitude - radius (as given by the \textsc{Source Extractor} `\textsc{radius}' parameter) parameter spaces, where the galaxy, star and artefact locus are well separated.

\paragraph*{\textit{Shape measurement and PSF correction}}
The weighted second order ($M_{xx}$, $M_{xy}$, $M_{yy}$) and fourth order ($M_{xxxx}$, $M_{xxxy}$, $M_{xxyy}$, $M_{xyyy}$, $M_{yyyy}$) moments of all deblended sources are first measured on the coadded F814W image. To correct these measurements for the detector response, the PSF and its variations across the field of view are then modelled by comparing the measured star moments with \textsc{Tiny Tim} PSF models \citep{krist2011}, fitting the focus parameter for each individual exposure. The resulting PSFs are then interpolated at the positions of the galaxies, combined and rotated to the reference frame of the drizzled image, to obtain a "coadded" PSF model, which is used to correct the galaxy moments. The stars used for the PSF modelling are selected as described above, resulting in between 70 and 200 stars for each field, with magnitudes in the ACS/F814W band in the range 19-23.

The galaxy ellipticities are derived as:
\begin{align}    
    \chi_1 &= \frac{M_{xx}^{\mathrm{corr}} - M{yy}^{\mathrm{corr}}}{M_{xx}^{\mathrm{corr}} + M_{yy}^{\mathrm{corr}}},\\
    \chi_2 &= \frac{2M_{xy}^{\mathrm{corr}}}{M_{xx}^{\mathrm{corr}} + M_{yy}^{\mathrm{corr}}},
\end{align}
where $M_X^{\mathrm{corr}}$ are the second order PSF-corrected moments for the galaxies.
The shear is then obtained from the mean ellipticity as:
\begin{equation}
    g = C\frac{\langle\chi\rangle}{G},
\end{equation}
where $C$ is the calibration constant fixed to $C^{-1}=0.86$, as measured from simulations in \citet{leauthaud2007} to correct for the multiplicative bias, and the polarisability $G$ is computed from the second and fourth order moment as shown in Equations 15 to 17 in \citet{harvey2019}.

Some additional steps are finally performed to clean the catalogue: (1) polygonal masks are applied around saturated stars to remove sources potentially contaminated by star spikes; (2) double detections of a single source are removed. This produces the baseline pyRRG source catalogue, to which we apply additional cuts as described in the following section, to ensure maximum purity, and select only galaxies that are located behind the lensing cluster and therefore carry the weak lensing signal.

    \subsection{Catalogue clean and background source selection}
    \label{sec:cuts}

\paragraph*{\textit{Catalogue cleaning.}} To further remove contaminants in the catalogue, we perform some additional cuts:
\begin{itemize}
    \item sources located close to the edges of the different exposures, which could potentially affect their shape measurements, by applying a threshold on the minimum number of exposures for each detected source, which removes $1-5\%$ of sources;  
    \item objects smaller than the PSF, i.e. with a size smaller than $0.11\arcsec$; 
    \item very faint sources, with a signal-to-noise ratio lower than 7, defined as $\rm{SNR} = \frac{\rm{FLUX\_AUTO}}{\rm{FLUXERR\_AUTO}}$; 
    \item very bright sources, with a cut at 23.5 mag in ACS/F814W in all fields; 
    \item galaxies with ill-defined moments, i.e. with shear $\gamma < 0$ or $> 1$, or with a negative error.
\end{itemize}

\paragraph*{\textit{Background source selection.}}
Finally, to obtain the weak lensing catalogue for each cluster, we need to only select galaxies located behind the lens, as the images of foreground and cluster-member galaxies are not distorted by the gravitational field of the cluster, and would therefore dilute any measurement. This step would be relatively straightforward if all galaxies in the field had a measured redshift, but this is not the case. For the fraction of galaxies that do have a spectroscopic redshift (see Sect.~\ref{sec:redshifts}), we simply remove all galaxies that have a redshift $z < z_{\rm{cl}} + \delta z$, where $z_{\rm{cl}}$ is the redshift of the cluster, and $\delta z$ is fixed to $\sim 0.03$ for all six galaxy clusters of our sample. 

For the remaining sources that do not have a measured redshift, we apply calibrated selection cuts, based on their colours and magnitudes, as follows.
First, we extract the photometry in the F160W, F814W and F606W bands for all galaxies. For each band, we run \textsc{Source Extractor} in dual image mode, where the detection is always done in the F814W band. To further ensure that the source detection matches the one obtained in the galaxy detection and shape measurement pipeline (see Sect.~\ref{sec:pyrrg}), each run is performed in the hot/cold configuration, using the same \textsc{Source Extractor} parameters as in pyRRG. The three photometric catalogues are combined together into a multi-band one, which is then matched with the cleaned weak-lensing catalogue. 

\begin{figure}
    \begin{center}
    \hspace{-5mm}\includegraphics[width=0.5\textwidth,angle=0.0]{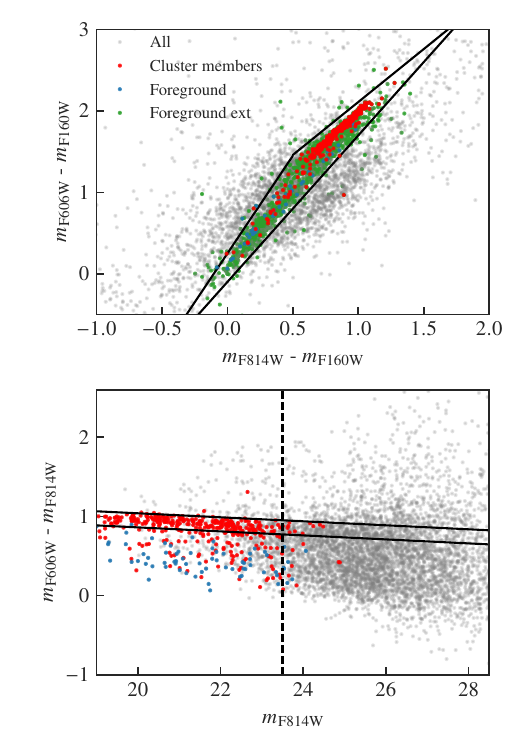}
    \caption{Background galaxy selections for the cluster AS1063. \emph{Top:} Colour-colour diagram ($m_{\mathrm{F814W}}$-$m_{\mathrm{F160W}}$) \emph{vs} ($m_{\mathrm{F606W}}$-$m_{\mathrm{F160W}}$) for objects with good WFC3/F160W, ACS/F814W and ACS/F606W photometry. Grey dots represent all galaxies with both ACS and WFC3 imaging. Non-lensed galaxies, diluting the shear signal, are marked by different colours: galaxies identified as foreground with spectroscopic redshifts $z < 0.34$ (blue), and galaxies classified as cluster members due to their spectroscopic redshifts $0.34 < z < 0.4$ (red). As only a small number of foreground contaminants are identified in each field, to better calibrate the colour-colour selection we show in addition all galaxies with $z < 0.4$ in the six BUFFALO fields and in the Candels dataset (`Foreground ext', in green).  The solid black lines delineate the colour-colour-cut defined for this work to mitigate shear dilution by non-lensed galaxies. 
        \emph{Bottom:} Colour-magnitude diagram $m_{\mathrm{F814W}}$ \emph{vs} ($m_{\mathrm{F606W}}-m_{\mathrm{F814W}}$) for galaxies within the ACS field without WFC3 imaging. The same colour code as above is applied. The bright source cut is shown as a dashed horizontal line. }
    \label{fig:ccselec}
    \end{center}
\end{figure}

As the WFC3/F160W observations cover a smaller region of the clusters than the two ACS bands, a fraction of the galaxies are only observed in the ACS/F814W and ACS/F606W bands. In addition, a small fraction of galaxies in each cluster field is only detected in the ACS/F814W footprint. The number of galaxies in each case is given in Table~\ref{tab:source_stats}. We apply slightly different approaches for the three samples, following the method described in detail in e.g. \citet{jauzac2012} and \citet{niemiec2023}: 
\begin{itemize}
    \item For galaxies observed in the three photometric bands, we perform a selection in the colour-colour parameter space represented in the top panel of Fig.~\ref{fig:ccselec}, $m_{\rm F814W}$--$m_{\rm F606W}$--$m_{\rm F160W}$. To define the boundaries of the region in the colour-colour diagram that contains most of the foreground contaminants and should therefore be excluded from the catalogue, we use galaxies with a measured spectroscopic redshift. As the number of such galaxies is relatively low for each individual cluster, we combine data from the six clusters to perform that calibration. To further increase the statistics,  we include in the calibration dataset other galaxies observed with \emph{HST} in the same bands (WFC3/160W, ACS/F814W and ACS/F606W) and having spectroscopic redshift estimations. For this, we use the CANDELS data\footnote{\url{http://arcoiris.ucolick.org/candels/index.html}} \citep{grogin2011, koekemoer2011}, and in particular the COSMOS, EGS and GOODS-S fields.  
    The galaxies removed from our final catalogue are the ones located within the black triangle in Fig.~\ref{fig:ccselec}.
    \item For galaxies that are only in the F814W and F606W footprints, we apply a colour-magnitude selection, which is most appropriate to select elliptical, passive cluster member galaxies. This is called the red-sequence method, as these galaxies present a tight correlation between their colour and magnitude \citep{gladders2000}. We fit the red sequence in the $m_{\rm F606W}$--$m_{\rm F814W}$ vs $m_{\rm F814W}$ parameter space, using an iterative 3-sigma clipping process. We select as cluster members galaxies located within $1\sigma$ of the best-fit red sequence, as shown in the bottom panel of Fig.~\ref{fig:ccselec}. In addition, the bright galaxy magnitude cut at $m_{\rm F814W} = 23.5$ removes a fraction of the foreground galaxies.
    \item Finally, for the small fraction of galaxies that only have ACS/F814W photometry, we apply a simple magnitude cut, keeping as background galaxies only those with $m_{\rm F814W} > 26.5$.
\end{itemize} 
Finally, we remove galaxies located in the strong-lensing region of each cluster, a roughly elliptical region with radius $\sim 1'$ encompassing all the identified strong-lensing features, as their distortion may be too high to be treated in the weak lensing approximation. Although this removes the regions where the distortion is the highest, we examine in more detail if the approximation holds in the remainder of the BUFFALO footprint in Sect.~\ref{sec:wl_approx}. The final number of background sources, as well as the corresponding source density, is given in Table~\ref{tab:source_stats}. The depth of the space-based images provided by the BUFFALO program give rise to very high source density, about 50 sources/arcmin$^2$, as compared with stage-IV surveys, such as \emph{Euclid} and LSST, which forecast densities $\sim 30$ sources/arcmin$^2$.

 \begin{table*}
    \centering
    \begin{tabular}{c|c|c|c|c c c|c|>{\bfseries}c|c|c}
\multirow{2}{*}{Galaxy cluster}     & \multirow{2}{*}{Clean} & \multirow{2}{*}{$z_{\rm{spec}}$} & \multirow{2}{*}{$z_{\rm{phot}}$}  & \multicolumn{3}{c|}{No $z$}    & \multirow{2}{*}{Final}  & \multirow{2}{*}{$n_{\rm{sources}}$}   & \multirow{2}{*}{$<z>$} &   $\mathcal{A}^{\rm{eff}}$    \\
            &       &       &       & 3 bands   & 2 bands   & 1 band    &       &       &       &      \\
\hline
A2744       & 5503  & 104   & 2781  & 3715      & 1623      & 61        & 3631  & 62    & 1.73  & 58    \\
A370        & 4582  & 883   & 2131  & 1693      & 940       & 1066      & 2917  & 45    & 1.74  & 65    \\
AS1063      & 5180  & 99    & 2080  & 3501      & 1530      & 50        & 3288  & 52    & 1.70  & 59    \\
M0416       & 5561  & 203   & 2753  & 3528      & 1665      & 165       & 3708  & 66    & 1.65  & 56    \\
M0717       & 5282  & 38    & 2707  & 3510      & 1566      & 168       & 2944  & 53    & 1.71  & 55    \\
M1149       & 5712  & 125   & 2874  & 3762      & 1741      & 84        & 3660  & 69    & 1.89  & 53    \\
    \end{tabular}
    \caption{Weak lensing source statistics for the six lens cluster fields, from left to right: name of the cluster, number of sources in the catalogue after the cleaning cuts; number of sources with a spectroscopic redshift; number of sources without a spectroscopic redshift but with a photometric redshift; number of sources without a spectroscopic redshift measurement but with magnitudes measured in either 3, 2 or 1 photometric bands; final number of background galaxies in the catalogue,  source density in arcmin$^{-2}$, mean weighted redshift of the sample, and  effective, unmasked area in arcmin$^2$.}
    \label{tab:source_stats}
\end{table*}

    \subsection{Redshifts}
    \label{sec:redshifts}
    
Redshifts are an essential piece of information when producing and using a weak lensing catalogue. First, as described in Sect.~\ref{sec:cuts}, they are useful to remove foreground and cluster-member galaxies that, otherwise, would dilute the lensing signal. Second, as the lensing effect is sensitive to the observer-lens-source geometry, they are needed to properly measure the signal, and resulting cluster total mass (see Sect.~\ref{sec:dsigma}).
In our catalogue, three cases are encountered: galaxies that have (1) a spectroscopic redshift, (2) a photometric redshift, or (3) none of the above. We summarise here the data used as well as the implications for each  scenario. In Table~\ref{tab:source_stats}, we also present the number of galaxies in each category, for each of the six clusters.

\paragraph*{\textit{Spectroscopic redshifts.}}
Spectroscopic coverage of the BUFFALO fields is available throughout the HST/BUFFALO footprint, drawn from a variety of publicly-available sources. A majority of redshifts are extracted from deep MUSE mosaics taken as part of the MUSE GTO lensing clusters program \citep[PI J. Richard;][]{mahler2018, lagattuta2019} as well as some early data taken during MUSE commissioning. These redshifts are supplemented by grism spectroscopy taken from GLASS \citep[the Grism Lens-Amplified Survey from Space; PI T. Treu;]{Schmidt2014} as well as other archival coverage available in the NASA Extragalactic Database (NED). The complete redshift catalogue, combines information from each individual source (RA, Dec, redshift, and confidence value) into a master list. To do this, we first visually inspect the positions of each entry by overlaying their coordinates on the BUFFALO images (which are themselves aligned to the GAIA DR2 WCS) and adjust the centroid values of each object to match the
observed \emph{HST} position. Next, we normalize the confidence values of each redshift to a common scale, choosing the 3 (secure redshift) / 2 (highly probable redshift) / 1 (low confidence/guess redshift) convention employed by the MUSE GTO team \citep[e.g.][]{lagattuta2019, Richard2021}. Finally, we remove any duplicate entries that appear in multiple catalogues, leaving a single redshift entry for each distinct object. In cases where overlapping entries disagree on the final redshift, we select the entry with the higher confidence value. Combined, the final catalogues contain 1452 redshifts over all 6 clusters, which we then use for cluster member identification, foreground contamination removal, and lens modelling.

\paragraph*{\textit{Photometric redshifts.}}
We match galaxies without a spectroscopic redshift with photometric catalogues produced with \textit{HST}, \textit{K}-band and Spitzer observations as described in \citet{pagul2024}, that contain photometric redshifts estimates from the template fitting software \textsc{lephare} \citep{arnouts1999, ilbert2006}. As shown in Table~\ref{tab:source_stats}, about half of the weak-lensing sources are included in these photometric catalogues. This is mostly because photometric catalogues only cover the regions that have been observed with both ACS and WFC3: in these regions, about 80\% of the weak lensing sources have a counterpart in the photometric catalogues. The remaining difference could probably be attributed to the different parameters used for the source extraction, as the weak lensing pipeline is specifically tuned to select small background sources.
We verify the precision of the photometric redshifts by comparing them with the spectroscopic ones for A370, for which we have the largest number of spectroscopic redshifts. We first select photometric redshifts with the highest confidence measurement, choosing $\chi^2_z < 2$ as the threshold, and match them with the full spectroscopic catalogue of this cluster. It results in 563 galaxies with both measurements. Out of this sample, $15\%$ present ``catastrophic failures'', which we define as $|z_{\rm{phot}} - z_{\rm{spec}}| > 0.5(1 + z_{\rm{spec}})$. For the remaining $85\%$, the RMS deviation is $\sigma_z = 0.28$, which suggests that the scatter in the photometric redshift estimates is too large to use them to properly select background sources from cluster members and foreground contaminants. 
We verify this by comparing the lensing signal obtained by selecting the background sources either using only the colour-colour and colour-magnitude diagrams as described in Sect.~\ref{sec:cuts}, either using photometric redshifts (and colour diagram for galaxies without a photometric estimate). Indeed, using photometric redshifts induces a higher contamination rate in the background source catalogue, and therefore dilutes the lensing signal. This results in systematically underestimated cluster masses as compared to measurements obtained with only colour-colour selections, by approximately 0.5~dex.
On the other hand, even if the scatter in the photometric estimates is high, the overall distribution does not present a significant bias. We therefore still use the photometric redshift measurements to derive the ensemble redshift distributions for the sources, as described in the next paragraph. This may cause some deviations in the $N(z)$ as compared to the true source distribution, but we show in Sect.~\ref{sec:Nzsyst} that it should not impact our results.

\paragraph*{\textit{Redshift distribution function $N(z)$.}}
For each cluster, the large majority of background galaxies are left without a reliable measurement for their redshift. As described in Sect.~\ref{sec:cuts}, we remove most of the foreground and cluster-member galaxies from these samples by applying calibrated colour and magnitude cuts. However, we still need a statistical estimate of their redshift distribution, $N(z)$, to compute the cluster lensing signal (see Sect.~\ref{sec:dsigma}). This should account both for the intrinsic evolution of the number of galaxies over time, and for the selection function of the observations and particularities of the field. For instance, background galaxy catalogues are selected to remove most of the sources located at or below the redshift of the corresponding cluster lens. Due to the magnification of the cluster lens, they could also contain a different ratio of high-redshift sources than an equivalent field galaxy sample observed at the same depth.

\begin{figure}
    \begin{center}
    \hspace{-5mm}\includegraphics[width=0.5\textwidth,angle=0.0]{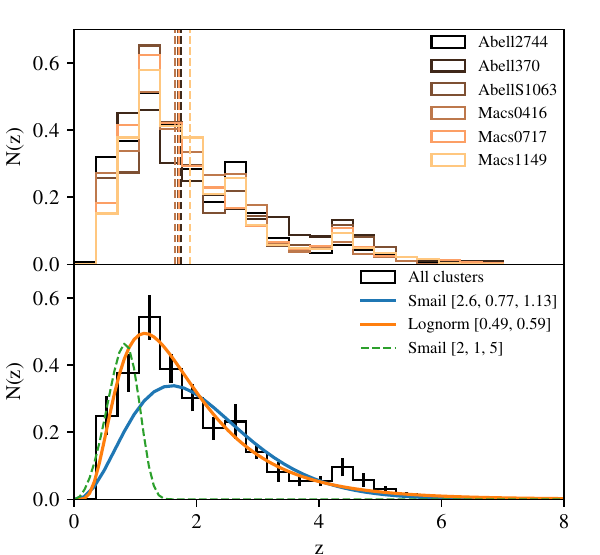}
    \caption{Redshift distribution function $N(z)$ for the background galaxy population. \emph{Top panel:} redshift distribution for the background source distribution for each cluster, estimated from galaxies with either a measured spectroscopic or photometric redshift. Dashed vertical lines represent the weighted mean of the redshift distribution for each cluster. \emph{Bottom panel:} the black histogram represents the $N(z)$ for all six galaxy clusters combined, where  error bars show the standard deviation between clusters. The blue curve represents the fit of a \citet{smail1995} distribution to the histogram, and the orange curve of a log-normal distribution. In both cases, the best-fit parameters are given in the legend between brackets. The dashed green curve represents the \citet{smail1995} distribution with the original parameters, also given in brackets. We note the small but statistically significant peak at $z\sim 4.3$, which could be due to a systematic bias in the photometric-redshift estimation, but would require further investigations. }
    \label{fig:nz}
    \end{center}
\end{figure}

To best account for the combination of these different effects, we estimate the $N(z)$ distribution empirically: for each galaxy cluster, we select only galaxies with a spectroscopic or photometric redshift, to which we apply the different selection steps.  In this way, by passing the sample through the different colour and magnitude cuts, the $N(z)$ can account for the residual contamination of the catalogues by foreground and cluster member galaxies. Top panel of Fig.~\ref{fig:nz} presents the redshift distribution of the background sources for each of the lens clusters. Although the exact values of the selection thresholds vary to account for the different redshifts of the clusters, the redshift distributions are generally consistent with each other. We therefore choose to measure the $N(z)$ globally, by combining all histograms together. This could include some bias in our weak lensing measurements, for instance, by overestimating the number of low redshift contaminants in the analysis of the higher-redshifts clusters, and we further examine this effect in Sect.~\ref{sec:Nzsyst}.

The bottom panel of Fig.~\ref{fig:nz} presents the combined distribution of the background galaxy redshifts for all six clusters as a black histogram. The associated error bars represent the standard deviation among the clusters. To use the $N(z)$ in the weak-lensing analysis, it is more convenient to work with an analytical function. We, therefore, fit the combined histogram with two different analytical distributions. First, we use the distribution proposed in \citet{smail1995}:
\begin{equation}
    N(z)^{\rm{Smail}} = z^{p_1}\exp{-\left(\frac{z}{p_2}\right)^{p_3}},
    \label{equ:smail95}
\end{equation}
where $p_1$, $p_2$ and $p_3$ are free parameters. The best-fit distribution is shown in blue in the bottom panel of Fig.~\ref{fig:nz}, corresponding to parameters $p_1 = 2.6 \pm 0.3$, $p_2 = 0.77 \pm 0.08$ and $p_3 = 1.13 \pm 0.07$. However, this function does not account well for the distribution at low redshift, and we examine and alternative fit, using a log-normal distribution defined as follows:
\begin{equation}
    N(z)^{\rm{lnorm}} = \frac{1}{z\sqrt{2\pi\sigma^2}}\exp{\left( -\frac{(\ln{z} - \mu)^2}{2\sigma^2} \right)},
    \label{equ:lognorm}
\end{equation}
with free parameters $\mu$ and $\sigma$. The best-fit log-normal distribution is shown in orange in the bottom panel of Fig.~\ref{fig:nz}, where the associated best-fit parameters are $\mu = 0.49 \pm 0.03$ and $\sigma = 0.59 \pm 0.02$. This function represents well the variations of the distribution at all redshifts, and we keep it as the baseline $N(z)$. We test how the use of the different redshift distributions impact the measured lensing signal in Sect.~\ref{sec:Nzsyst}.

\section{Weak lensing mass of the BUFFALO clusters}
\label{sec:wl_masses}

In this section, we present the methodology used to derive the fiducial weak-lensing masses for the clusters, as well as for the strong+weak lensing masses.

    \subsection{$\Delta\Sigma$ measurement}
    \label{sec:dsigma}

The quantity that is measured, and which is more straightforward to physically interpret than shear, is the excess surface mass density profile, $\Delta\Sigma (R)$, in comoving units, as a function of the distance to the cluster centre. It is related to the tangential shear, $\gamma_t$, as:
\begin{equation}
    \Delta\Sigma (R) = \Sigma_{\rm{crit}}\gamma_t = \bar{\Sigma}(< R) - \bar{\Sigma} (R),
\end{equation}
where $\bar{\Sigma}(< R)$ is the mean surface density in a disk of radius $R$ and $\bar{\Sigma} (R)$ the density at the radius $R$. $\Sigma_{\rm{crit}}$ is the critical surface density in comoving units, which is a function of the angular diameter distance to the lens, $D_{\rm{l}}$, to the source, $D_{\rm{s}}$, and between the lens and the source, $D_{\rm{ls}}$, and is defined as:
\begin{equation}
    \Sigma_{\rm{crit}} = \frac{c^2}{4\pi G}\frac{D_{\rm{s}}}{D_{\rm{l}}D_{\rm{ls}}}\frac{1}{(1 + z_{\rm{cl}})^2}.
\end{equation}
$c$ represents the speed of light, $G$, the gravitational constant and $z_{\rm{cl}}$, the redshift of the lens. The critical surface density indicates the limit where lensing locally becomes strong (where $\Sigma \gtrsim \Sigma_{\rm crit}$), and its value varies from $\sim 2000\,M_{\odot}/{\rm pc}^2$ to $\sim 3000\,M_{\odot}/{\rm pc}^2$ for the six clusters.

In practice, we can only access the \emph{reduced} shear from observations, defined as $g_t = \frac{\gamma_t}{1 - \kappa}$, and therefore measure the \emph{reduced} excess surface mass density profile, that we note $\Delta\Sigma_r$ and define as :
\begin{equation}
\Delta\Sigma_r = \Sigma_{\rm crit}g_t = \Sigma_{\rm crit}\frac{\gamma_t}{1 - \kappa}.
\end{equation}
We estimate $\Delta\Sigma_r$ by adding the signal from sources split in 10 radial logarithmic bins, from 0.2 to 5\,cMpc$/h$ (where the `c' in front of Mpc indicates the use of comoving units). We chose as the cluster centre the reference positions given in Table~\ref{tab:clusters}, but we examine in Sect.~\ref{sec:miscentring} the impact of alternative centre choices. To account for shape measurement errors and the different signal-to-noise of each source, we weigh the measured signal with an inverse weight factor associated with each source galaxy, $w_s$. For each cluster, $\Delta\Sigma$ is then estimated as:
\begin{equation}
    \Delta\Sigma_r (R) = \frac{\sum_{\rm{s}}{ W_{\rm{s}} g_t^{\rm{s}} \Sigma_{\rm{crit}}(z_{\rm{s}}, z_{\rm{cl}})}}{\sum_{\rm{s}}{ W_{\rm{s}}}},
\end{equation}
where $W_{\rm{s}} = w_{\rm{s}}/(\Sigma_{\rm{crit}}(z_{\rm{s}}, z_{\rm{cl}}))^2$, and $w_{\rm s}$ is the inverse-variance weight computed for each source. The sum is performed over all sources in each bin, with $z_{\rm{s}} > z_{\rm{cl}} + z_{\rm{err}}$ to ensure that no sources located in the foreground of the cluster are being used. This lensing signal is computed for each cluster using a modified version of the \textsc{athena}\footnote{\url{http://www.cosmostat.org/software/athena/}} software, a 2d-tree code estimating second-order correlation functions.

\begin{figure*}
    \begin{center}
    \hspace{-5mm}\includegraphics[width=\textwidth,angle=0.0]{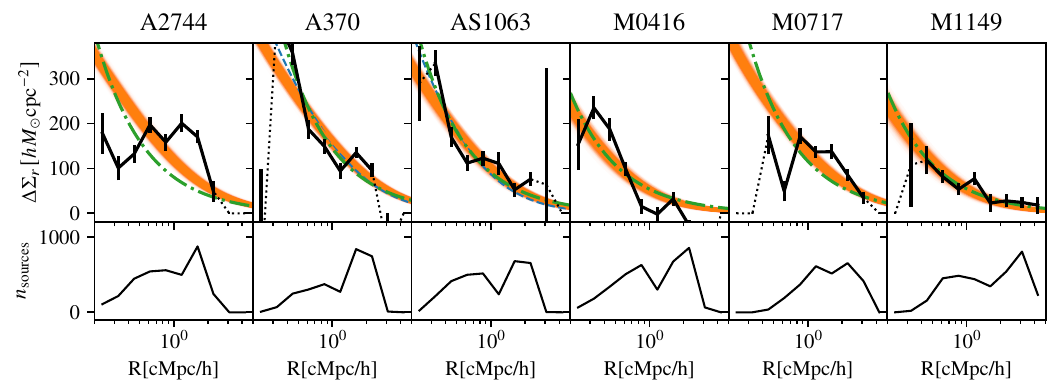}
    \caption{\emph{Top panel:} in black, $\Delta\Sigma_r$ profiles for each of the six galaxy clusters. The error bars account for the uncertainty coming from the missing redshift information for a fraction of the sources, as well as the statistical errors. Only bins where the number of background sources is higher than a given threshold are used to fit the model (solid line). For each cluster, we represent the best-fit NFW model with the \citetalias{darragh-ford2023} $M-c$ relation  as a solid orange line, and 1000 models corresponding to randomly drawn MCMC samples to show the dispersion. In addition, for A370 and AS1063 the ``Flat $c$ prior'' best-fit models are shown with blue dashed lines. Finally, the green dot-dashed lines correspond to the Singular Isothermal Sphere (SIS) models described in Sect.~\ref{sec:SIS}.  \emph{Bottom panel:} number of background sources used to compute the lensing signal in each radial bin. The y-axis is in linear scale. }
    \label{fig:DSig}
    \end{center}
\end{figure*}

As discussed in Sect.~\ref{sec:redshifts}, some background galaxies do not have a measured redshift, necessary to compute $\Delta\Sigma_r$. For this population, we therefore randomly draw a value for their redshift, from the $N(z)$ distribution measured as described in Sect.~\ref{sec:redshifts}. To account for the error associated with this random redshift assignment, we perform a hundred realisations of this random sampling, and compute the lensing signal for each realisation of the background source catalogue. For a given cluster, the lensing signal is computed as the mean over these 100 measurements.
Moreover, to take into account statistical uncertainties, we perform a bootstrap on the source catalogues, meaning that for each random redshift realisation, we create 50 subsampled source catalogues, each containing 80\% randomly selected of the initial catalogue. We measure $\Delta\Sigma_r$ for each  subsampled source catalogues. The error on the lensing signal is then computed as the standard deviation over the $100 \times 50$ signals measured. The $\Delta\Sigma$ profile and associated error for each cluster is presented in the top panels of Fig.~\ref{fig:DSig}. We note that we consider the covariance matrices to be diagonal: as for each cluster there is no overlap in the sources included in each radial bin, we expect the covariances between bins to be negligible. We verify this assumption in Appendix~\ref{sec:cov}. Given the small radial range of the considered measurement, we also neglect the contribution of the large-scale structure (LSS) covariance \citep[see e.g.][]{wu2019}.
Finally, we show in the bottom panels of Fig.~\ref{fig:DSig} the number of background sources used to compute the signal in each radial bin. Because of the geometry of our observations, the number of sources does not increase monotonically with radius, and presents a dip at the limit between the two BUFFALO fields. To ensure good statistics, we do not use in the fitting procedures the signal measured in bins where the number of background sources is lower than 5\% of the maximum number of sources in any other bin for the same cluster. Discarded parts of the $\Delta\Sigma_r$ profiles are shown as dotted lines in the top panel of Fig.~\ref{fig:DSig}.

    \subsection{Model fitting and cluster masses}
    \label{sec:model_res}

To estimate cluster total masses from the measured weak lensing signal, we fit a simple model to each of the $\Delta\Sigma_r$ profiles. We consider that each cluster can be well described by a single halo, centred on the same position as the centre of the azimuthally-averaged profile that we computed (the coordinates are given in Table~\ref{tab:clusters}). This assumption is commonly made in stacked weak-lensing analyses of clusters, and generally holds well when considering relaxed, unimodal clusters. However, this is not necessarily a good description of the clusters in our sample: they are very complex clusters, some of them composed of massive sub-clusters in the process of merging (see for instance \citealt{limousin2016} for M0717; \citet{Bergamini2023b} for A2744; \citet{lagattuta2019, niemiec2023} for A370; \citet{Grillo2016} for M1149; \citet{Rihtarsic2025} for M0416). We use this simple 1-halo model to derive a baseline estimate of the cluster mass, and test the impact of the different simplifying assumptions in Sect.~\ref{sec:discussion}. 

The distribution of matter within the halo is described with a NFW mass density profile \citep{nfw1996}:
\begin{equation}
    \rho (r) = \frac{\delta_c \rho_{\rm{crit}}(z_{\rm{cl}})}{(r/r_{\rm{s}})(1 + r/r_{\rm{s}})^2},
\end{equation}
where the two characteristic radii of the profile, the scale radius, $r_{\rm{s}}$, and the virial radius, $r_{200}$, are related through the dimensionless concentration by $r_{\rm{s}} = r_{200}/c$. The virial radius is defined as enclosing a spherical region where the density is 200 times higher than the critical density of the Universe at the redshift of the cluster, $\rho_{\rm{crit}} (z_{\rm{cl}})$, which gives:
\begin{equation}
    \frac{M_{200}}{4/3\pi r_{200}^3} = 200\rho_{\rm{crit}}(z_{\rm{cl}}).
\end{equation}
Finally, $\delta_c$ represents the characteristic overdensity for this halo definition, and is a function of the concentration, as:
\begin{equation}
    \delta_c = \frac{200}{3}\frac{c^3}{\ln(1 + c) - c/(1 + c)}.
\end{equation}
In summary, the NFW mass density profile can be expressed as a function of two free parameters, the virial mass, $M_{200}$, and the concentration parameter, $c$, as $\rho_{\rm{NFW}}(r | M_{200}, c)$. By integrating the 3D mass density profile along the line-of-sight, an analytical expression for $\Sigma_{\rm{NFW}}(R)$ can be derived \citep[see Equation 11 in][]{wright2000}, which can, in turn, be radially integrated to obtain $\bar{\Sigma}(< R)$ \citep[see Equation 13 in ][]{wright2000}.
Finally, as our measured signal is the \emph{reduced} excess surface mass density, we define the model for $\Delta\Sigma_r$ as \citet[see e.g][]{sereno2025}:
\begin{equation}
\Delta\Sigma_r(R) = \frac{\Delta\Sigma}{1 - \Sigma_{\rm NFW}(r)/\Sigma_{\rm crit}},
\end{equation}
as the convergence $\kappa$ that transforms the shear to the reduced shear is $\kappa = \Sigma/\Sigma_{\rm crit}$.
The resulting model $\Delta\Sigma_r(R | M_{200}, c)$ has two free parameters, and depends on the cluster redshift through the critical density of the Universe $\rho_{\rm crit}$, and on the source redshifts through the critical density $\Sigma_{\rm crit}$. For the later, we use the mean redshift of each background source sample, given in Table~\ref{tab:source_stats}.

We obtain best-fitting parameters and credible intervals through a Markov Chain Monte Carlo (MCMC) sampling, in the python \textsc{emcee} implementation \citep{foreman-mackey2013}. The likelihood describing our model is expressed as:
\begin{equation}
    \mathcal{L} = \prod_{i=1}^{N} \frac{1}{\sqrt{2\pi\sigma^2_i}}\exp\left(-\frac{1}{2\sigma_i^2}(\Delta\Sigma_r^{\rm{obs}}(R_i) - \Delta\Sigma_r^{\rm{mod}}(R_i | M_{200},c))^2\right),
\end{equation}
where the product is carried out over the $N=10$ radial bins. We assume flat and broad priors for the two free parameters, $13.5 < \log{M_{200}} < 15.5$ and $2 < c < 18$\footnote{Alternatively, we tested a concentration prior uniform in log-space ($0 < \log{c} <1$), but found no significant difference.}.
To sample the posterior distribution, we run 10 chains with 10 000 iterations each, and discard the first 1000 as a burn-in phase. We verify that the sampling is run for enough iterations by computing the mean autocorrelation time $\tau$, and verify that for each case $\tau \gg N_{\rm iter}$.

The chains converge well only for two clusters, A370 and AS1063. For the remaining four, the posterior distribution of the concentration is prior dominated, probably due to the small radial extent of our weak lensing constraints. However, as the concentration of a halo correlates with its mass, we can reduce the number of free parameters to 1 by introducing a mass-concentration ($M-c$) relation in the model, allowing us to derive a concentration value for each $M_{200}$ sample. To test the impact of this assumption, we re-run the optimisation with two different $M-c$ relations: (1) from \citet{darragh-ford2023} \citepalias[hereafter][]{darragh-ford2023}, measured from strong+weak lensing data on the CLASH clusters \citep{merten2015}, and (2) from \citet{duffy2008}, measured from N-body simulations, on their full sample of haloes, at redshifts $0 < z < 2$. The former relation should match best our observations, as the CLASH clusters are massive strong-lensing clusters, similar to the sample studied in this analysis. In particular, strong-lensing clusters could present higher concentrations than the general cluster population \citep[e.g][]{giocoli2012a}, and using a general $M-c$ relation such as the one from \citet{duffy2008} could bias the derived masses. We verify these assumptions on the two clusters A370 and AS1063. 

In Table~\ref{tab:results}, we present the best-fit parameters and the 68\% credible intervals for the three cases. For the run where both $M_{200}$ and $c$ are left free to vary, with flat priors for both (named ``Flat $c$ prior'' in Table~\ref{tab:results}),  results  not well converged are shown in light grey. For the two well converged clusters, A370 and AS1063, masses are consistent within 2$\sigma$ with the ones obtained using the \citetalias{darragh-ford2023} $M-c$ relation. 
For consistency, we consider as our baseline measurement the \citetalias{darragh-ford2023} values for all six clusters. In Fig.~\ref{fig:DSig}, we show the best-fit models with this $M-c$ relation for each cluster as a solid orange line, and represent the dispersion by drawing 1000 random samples and plotting the corresponding models. In addition, for A370 and AS1063, we show the ``Flat $c$ prior'' best-fit model for comparison with a dashed blue line.

\begingroup
\renewcommand{\arraystretch}{1.5} 

 \begin{table*}
    \centering
    \begin{tabular}{c | c c | c c | c c | c c | cc }
        & \multicolumn{2}{c|}{Flat $c$ prior}                               & \multicolumn{2}{c|}{\citetalias{darragh-ford2023}} & \multicolumn{2}{c |}{\citet{duffy2008}} & \multicolumn{2}{c |}{SL+WL Xcen} &  \multicolumn{2}{c}{SL+WL C0} \\
\hline
Cluster & $\log{M_{200}}$                   & $c$                           & $\log{M_{200}}$           & $c$               & $\log{M_{200}}$           & $c$       & $\log{M_{200}}$                  & $c$                    & $\log{M_{200}}$                  & $c$                    \\
\hline 
A2744   & \deemph{$15.28^{+0.04}_{-0.04}$}  & \deemph{$2.0^{+0.1}_{-0.1}$}  & $15.05_{-0.05}^{+0.05}$   & [3.7]             & $15.11_{-0.04}^{+0.04}$   & [2.9]     & $14.92_{-0.05}^{+0.05}$           & $2.1^{+0.1}_{-0.1}$   & $14.86_{-0.04}^{+0.04}$   & $6.0_{-0.3}^{+0.3}$  \\
A370    & $15.22_{-0.07}^{+0.08}$           & $4.1^{+1.0}_{-0.8}$           & $15.30_{-0.04}^{+0.04}$   & [3.1]             & $15.34_{-0.04}^{+0.04}$   & [2.7]     & $15.71_{-0.02}^{+0.02}$           & $2.7_{-0.5}^{+0.5}$    & $15.55_{-0.03}^{+0.03}$   & $2.5_{-0.1}^{+0.1}$  \\
AS1063  & $14.78_{-0.11}^{+0.12}$           & $6.4_{-2.0}^{+3.6}$           & $14.92_{-0.05}^{+0.05}$   & [4.1]             & $15.04_{-0.05}^{+0.05}$   & [2.9]     & $15.03_{-0.03}^{+0.03}$           & $4.8_{-0.2}^{+0.2}$   & $14.98_{-0.03}^{+0.03}$   & $5.3_{-0.2}^{+0.2}$  \\
M0416   & \deemph{$14.34_{-0.05}^{+0.06}$}  & \deemph{$15.5_{-3.7}^{+3.1}$} & $14.49_{-0.06}^{+0.06}$   & [5.6]             & $14.60_{-0.07}^{+0.07}$   & [3.1]     & $14.51_{-0.06}^{+0.06}$           & $5.7_{-0.5}^{+0.6}$   & $14.48_{-0.06}^{+0.06}$   & $6.2_{-0.6}^{+0.6}$ \\
M0717   & \deemph{$15.30_{-0.05}^{+0.05}$}  & \deemph{$2.1_{-0.1}^{+0.2}$}  & $15.21_{-0.05}^{+0.05}$   & [3.4]             & $15.26_{-0.05}^{+0.04}$   & [2.6]     & $15.19_{-0.05}^{+0.05}$           & $2.1_{-0.1}^{+0.1}$   & $15.35_{-0.07}^{+0.07}$   & $2.3_{-0.3}^{+0.3}$   \\
M1149   & \deemph{$14.61_{-0.14}^{+0.12}$}  & \deemph{$3.5_{-1.2}^{+3.6}$}  & $14.54_{-0.08}^{+0.07}$   & [5.4]             & $14.66_{-0.08}^{+0.07}$   & [3.0]     & $14.71_{-0.09}^{+0.08}$           & $2.6_{-0.4}^{+0.6}$   & $14.47_{-0.06}^{+0.06}$   & $9.7_{-1.3}^{+1.4}$  \\
    
    \end{tabular}
    \caption{Best-fit parameters for the six BUFFALO galaxy clusters. The masses, $M_{200}$, are expressed as $\log M_{200}/M_{\odot}$. We present different fitting approaches: (1) the two free parameters $M_{200}$ and $c$ are sampled, starting with flat priors for both of them (``Flat $c$ prior''), and (2) a concentration-mass relation is used, leaving only one free parameter, $M_{200}$. We test the mass-concentration relations from \citetalias{darragh-ford2023} (their fit to the strong+weak lensing data for the CLASH clusters), and from \citet{duffy2008} (full sample, for NFW haloes with $\Delta$=200). For the ``Flat $c$ prior'' case, clusters for which the concentration value did not converge well are shown in light grey. The last columns corresponds to the model fitted to the combination of strong and weak lensing measurements, using either the X-ray peak as centre (Xcen), or one of the BCGs (C0).}
    \label{tab:results}
\end{table*} 
\endgroup

    \subsection{Adding strong-lensing information in the core}
    \label{sec:sl+wl}

The HFF and BUFFALO galaxy clusters are notable in that they have been extensively observed and studied, and detailed lens models of the total mass distribution in their cores have been developed with SL \citep[e.g.,][]{zitrin2013, jauzac2015a, Grillo2016, mahler2018, Richard2021, lagattuta2022, beauchesne2023, Bergamini2023b, Diego2024, Schuldt2024} and WL \citep[e.g.,][]{Medezinski2016, Umetsu2016, Abriola2024, Harvey2024} or through combined SL+WL analyses \citep[e.g.,][]{jauzac2012, jauzac2015a, jauzac2016c, niemiec2023}. In these central regions, the distortion of the image of background sources can no longer be considered in the weak lensing approximation, but is instead observed in the strong lensing regime. Unlike weak lensing, where the reduced shear provides an observable directly related to the mass distribution, strong lensing corresponds to a non-linear regime, requiring the mass distribution to be inferred from the observed positions of multiple images of the same background galaxy — a complex inverse problem. Currently, the most widely used softwares for modelling the total mass distribution of strong-lensing clusters, e.g., \textsc{Glafic} \citep{oguri2010}; \textsc{GLEE} \citep{Suyu2010, Suyu2012}; \textsc{Lenstool} \citep{jullo2007};  \textsc{LTM or Zitrin-analytic} \citep{zitrin2012}; and \textsc{WSLAP+} \citep{diego2005, diego2016}, demand substantial user input to develop a single model, which limits their scalability for statistical studies.

\begin{figure}
    \begin{center}
    \hspace{-5mm}\includegraphics[width=0.5\textwidth,angle=0.0]{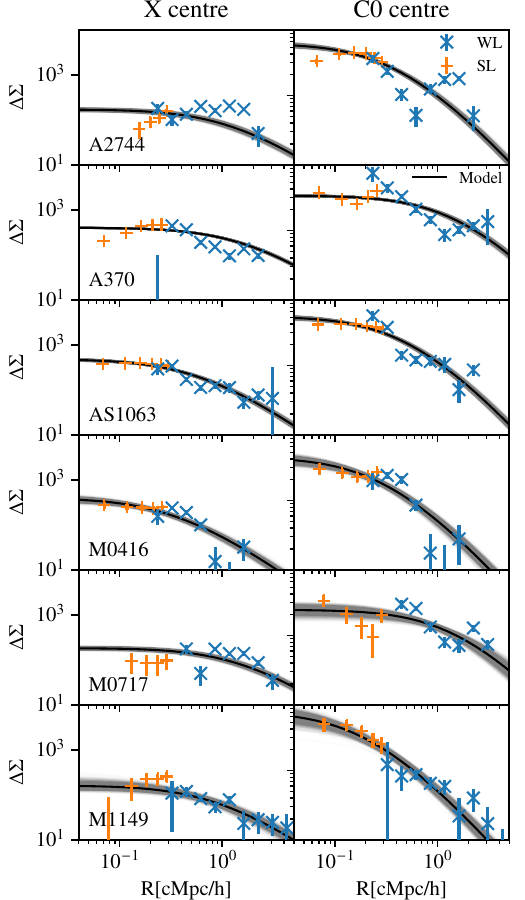}
    \caption{Excess surface density profiles for the six BUFFALO clusters, combining the weak lensing shear measurement at large radial scales (blue `x's) and strong-lensing mass models at small scales (orange `+'s). The best-fit model is shown as a black solid line, and 1000 models corresponding to randomly drawn MCMC samples are represented in grey to show the dispersion. $\Delta\Sigma$ is in $hM_{\odot}\rm{pc}^{-2}$. The profiles shown on the left panels are measured around the fiducial centres, i.e. X-ray brightness peaks, and around the brightest BCGs (named C0) on the right panels.}
    \label{fig:slwl}
    \end{center}
\end{figure}

However, we examine here the impact of this information at small radial scale when estimating the cluster mass from a 1D mass density profile. To this end, we use the projected mass distribution maps of the core of each cluster, which were obtained from detailed strong-lensing models using as constraints the observed positions of large samples of multiply-imaged background galaxies. For consistency, for all the clusters we use models which were produced with the \textsc{Lenstool} software, and published by Patel et al. (in prep.) for A2744; \citet{niemiec2023} for A370; \citet{beauchesne2023} for AS1063; \citet{Richard2021} for M0416; \citet{limousin2016} for M0717 and \citet{jauzac2016a} for M1149. From these maps, we compute the $\Delta\Sigma$ profiles by azimuthally averaging the mass in discs and annuli around the fiducial cluster centres, i.e. the X-ray brightness peaks. These profiles are shown as orange plus signs in Fig.~\ref{fig:slwl}, while the previously derived weak-lensing profiles are represented as blue `x' signs. We start by measuring profiles around the fiducial centres, i.e. the X-ray brightness peaks, shown on the left panels.

We fit a NFW profile on the full strong+weak lensing signal, which is shown as a black line on Fig.~\ref{fig:slwl}, and give the corresponding best-fit parameters in Table~\ref{tab:results}. The strong-lensing signal, measured at small radial scales, is very sensitive to the choice of the centre around which the signal is measured. We discuss the impact of this choice on the weak lensing signal in Sect.~\ref{sec:miscentring}, but we can already draw some conclusions from Fig.~\ref{fig:slwl}: for some galaxy clusters, AS1063 and M0416 in particular, the strong+weak lensing as measured around the X-ray peak seems to be well described by a NFW profile, while for the other four, there seems to be a strong decrease in the signal at small scales. This is due to the fact that in the latter case, the X-ray peak is not well aligned with a peak in the total mass distribution. We verify this by remeasuring the strong+weak lensing signal around one of the brightest cluster galaxies (BCGs), called C0, as the galaxy distribution is supposed to trace better the dark matter distribution, and BCGs are expected to be found at the (local) minima of the cluster gravitational potential. Positions of the X-ray peaks and BCGs are shown in the right panel of Fig.~\ref{fig:DSig_misc}, and the corresponding lensing signal measured around C0 in the right panels of Fig.~\ref{fig:slwl}. The best-fit NFW profile is shown as black curve on Fig.~\ref{fig:slwl}, and the corresponding parameters are given in Table~\ref{tab:results}. With this centre choice, all lensing profiles are better described with a NFW profiles, with some discrepancy still appearing, which can be largely explained as discussed in Sect.~\ref{sec:multimodality}, by the multimodal mass distribution of some of these complex clusters. We still consider this last mass measurement, called `SL+WL C0' as the most accurate, as it includes the largest number of observables spanning the largest spatial scales and seems to be least affected by model bias.

\section{Discussion}
\label{sec:discussion}

In this section, we test different assumptions  made when measuring the weak-lensing signal and modelling it, and verify if and how they impact the estimated cluster masses.

    \subsection{Impact of the choice of a centre}
    \label{sec:miscentring}

When measuring the weak-lensing signal produced by galaxy clusters, one important assumption is that of a centre, around which the shear is azimuthally averaged. This would usually corresponds to the BCG of the cluster or to the X-ray emission peak, and can be fairly straightforward to identify for relaxed clusters. However, for merging clusters such as the BUFFALOs, the notion of a centre can be more challenging to address: some clusters may have multiple BCGs, often corresponding to the respective centres of massive sub-haloes; in some cases, even if the BCG is easily identified, it may have been displaced from the minimum of the gravitational potential, and therefore be offset with respect to the true centre of the mass distribution. This can impact the measured weak-lensing mass, as this quantity is usually derived by fitting a spherically symmetric model, centred on a given point. In particular, this can be of importance in stacked cluster analyses, as they usually rely on automatic methods to identify clusters, and therefore their centres, which can lead to misidentifications. This effect is usually accounted for in the signal modelling, by adding a miscentring term in the theoretical $\Delta\Sigma$ \citep[see e.g.][]{simet2017}.

\begin{figure}
    \begin{center}
    \includegraphics[width=\linewidth,angle=0.0]{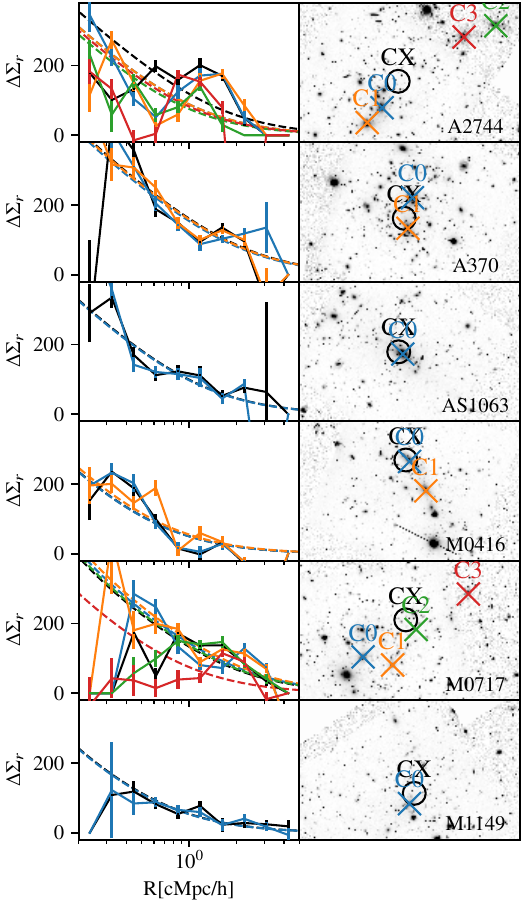}
    \caption{The impact of the cluster centre on our weak-lensing analyses. \emph{Left panel:} for each galaxy cluster, the lensing signal, $\Delta\Sigma$, measured around the different centres (solid line), and the best-fit NFW models (dashed line). The \citetalias{darragh-ford2023} $M - c$ relation was used to perform the fit. $\Delta\Sigma$ is in $hM_{\odot}\rm{pc}^{-2}$. \emph{Right panel:} positions of the centres, overlayed on the \textit{HST} F814W images. The size of each cutout is $4.75\arcmin \times 3.42\arcmin$.}
    \label{fig:DSig_misc}
    \end{center}
\end{figure}

Here, we quantify the impact of the centre choice on the measured lensing signal, and therefore on the mass measured for each cluster. As described before, we use as the fiducial centre the peaks in the X-ray brightness maps. To test the impact of this assumption, we identify for each cluster alternative centres, corresponding to the BCGs, shown on the right panels of Fig.~\ref{fig:DSig_misc}, overlaid on the \textit{HST} F814W images, where each image cutout is $4.75\arcmin \times 3.42\arcmin$. The different choices can be summarised as follows:
\begin{itemize}
    \item A2744 is composed of two sub-clusters, separated by $\sim 3\arcmin$ \citep[e.g.][]{Bergamini2023b, furtak2023}, the X-ray peak being located between these two haloes. The main, most massive component contains in turn two BCGs, that we name C0 and C1, separated by $d_{\rm{CX-C0}}=0.7\arcmin = 0.17\,{\rm{cMpc}}/h$ and $d_{\rm{CX-C1}}=1.1\arcmin = 0.27\,{\rm cMpc}/h$, respectively from the X-ray peak. The second component, located in the North-West of the cluster, also contains two BCGs, located at $d_{\rm{CX-C2}}=2.3\arcmin = 0.57\,{\rm cMpc}/h$ and $d_{\rm{CX-C3}}=1.6\arcmin = 0.40\,{\rm cMpc}/h$ from the X-ray peak. The multi-modality of the cluster mass distribution is well apparent on all lensing signals: the signal at small scales show  an important mass component located around each of the centres, while the secondary bump located at $\sim 1 - 2\arcmin$ is produced by the second, offset component. We examine how this bi-modality can be taken into account in the modelling of the signal in Sect.~\ref{sec:multimodality}.
    \item A370 presents a complex mass distribution, with two main BCGs and several massive substructures \citep[e.g.][]{niemiec2023}. The alternative centres are fixed at the positions of the two BCGs, C0 and C1, located at $d_{\rm{CX-C0}}=0.5\arcmin = 0.15\,{\rm cMpc}/h$ and $d_{\rm{CX-C1}}=0.2\arcmin = 0.06\,{\rm cMpc}/h$ from the fiducial centre. The centres are located relatively close to each other, and the different lensing signals show relatively monotonic behaviour.
    \item AS1063 shows the most relaxed morphology of the six clusters, as its mass distribution is uni-modal, but it is still the result of a recent merger, and presents an asymmetry in its X-ray emission \citep[e.g.][]{beauchesne2023}. The BCG, C0, is located at a small distance $d_{\rm{CX-C0}}=0.09\arcmin = 0.02\,{\rm cMpc}/h$ from the X-ray peak, and the resulting lensing profiles are well consistent.
     \item M0416 presents a bi-modal mass distribution, with a BCG in each component, C0 and C1 \citep[e.g.][]{bergamini2023}. They are separated from the fiducial centre by $d_{\rm{CX-C0}}=0.03\arcmin = 0.01\,{\rm cMpc}/h$ and $d_{\rm{CX-C1}}=0.7\arcmin = 0.22\,{\rm cMpc}/h$. Looking at the lensing profiles, it is not obvious if  one centre is more appropriate than the other, but it rather suggests a flatter and extended mass distribution.
    \item M0717 has a multi-modal mass distribution, with four main components identified in the cluster core \citep{limousin2016}. We use as alternative centres C0, C1, C2 and C3, the light peaks identified in each component by \citet{limousin2016} (in their analysis, C, D, B and A, respectively). The X-ray peak fiducial centre is located close to C2, between C0, C1 and C3 at separations $d_{\rm{CX-C0}}=1.1\arcmin = 0.45\,{\rm cMpc}/h$, $d_{\rm{CX-C1}}=0.9\arcmin = 0.37\,{\rm cMpc}/h$, $d_{\rm{CX-C2}}=0.3\arcmin = 0.12\,{\rm Mpc}/h$ and $d_{\rm{CX-C3}}=1.4\arcmin = 0.57\,{\rm cMpc}/h$ respectively.
    This cluster presents again, similarly to A2744, a main component around C0 and C1, and a secondary one at a radial separation of $\sim 2 -3\arcmin$, at the location of C3. The different lensing signals present a wide dispersion at small radial scales, but converge around $1-2\,\rm{Mpc}$.
    \item M1149 is elongated along the South-East/North axis \citep[e.g.][]{Schuldt2024}. Although the BCG (C0) can be identified quite clearly, the notion of centre is not quite clear for such a non-spherical object. The separation between the BCG and X-ray peak is $d_{\rm{CX-C0}}=0.2\arcmin = 0.08\,{\rm cMpc}/h$
\end{itemize}
For each measured weak-lensing signal, we fit an NFW profile following the procedure described in Sect.~\ref{sec:dsigma}, using the $M-c$ relation from \citetalias{darragh-ford2023}. The best-fit profiles are shown on the right panels of Fig.~\ref{fig:DSig_misc} as dashed lines, and the corresponding masses are given in Table~\ref{tab:masses_syst}. We also show all masses as green dots in the summary figure, i.e. Fig.~\ref{fig:summary}. For most clusters, the measurements are consistent within 1$\sigma$ with the fiducial measurement. Discrepancies arise for the most disrupted clusters, A2744 and M0717, where the strong multimodality of the mass distribution gives a higher weight to the centre choice. This could be studied more systematically on a larger cluster sample, by using some features that characterise the most unrelaxed clusters (e.g. distance between BCGs, magnitude gaps), and establishing a relation between the degree of relaxation of a cluster and the variation of the mass estimate with the choice of the centre.

    \subsection{Miscentring}
    \label{sec:random_miscentring}

\begin{figure}
    \begin{center}
    \includegraphics[width=\linewidth,angle=0.0]{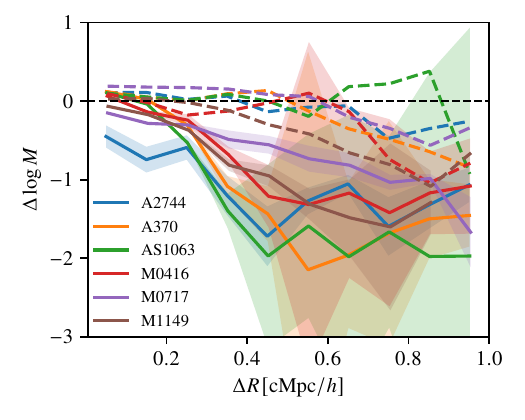}
    \caption{Mass bias when considering random choices for the centre of the shear profile, $\Delta\log M = \log M_{\rm{fiducial}} - \log M_{\rm{alt}}$, as a function of the distance between the centres, $\Delta R$. We remind the reader that the fiducial measurements are the ones performed around the X-ray brightness peaks, presented in Sect.~\ref{sec:dsigma}. The solid lines represent the mean over the different random realisations for each clusters, and the shaded area the dispersion. The dashed lines represent the remaining mass bias when considering an off-centred halo model. For clarity, the dispersion for the off-centred models is not represented, but it is equivalent or smaller than for the regular model.}
    \label{fig:DeltaMass}
    \end{center}
\end{figure}

In the previous section, we tested the impact of the cluster choice on the estimated mass, but using only centres that are associated to some real features of the cluster, i.e restricting to the ``sensible choices''. However, it is also possible to introduce a random error associated to the centre identification, for instance by choosing a line-of-sight galaxy which would not be associated to any feature in the cluster mass distribution. We test the impact of this kind of random error on our mass estimates. To this end, for each cluster, we draw 500 random centre positions, where the offsets between random centres and the fiducial one (i.e the X-ray emission peak) follow the distribution from \citet{zhang2019}, and the angle position is random.
For each random centre, we re-measure the lensing signal. We note that, to reduce computational costs, for each centre choice we compute the signal for one source catalogue only, assigning to each galaxy the mean redshift over the whole sample, as we have shown that this choice does not bias significantly the recovered signal (see Sect.~\ref{sec:Nzsyst}). For each signal, we fit an NFW profile to recover the cluster mass, $M_{200}$, assuming the \citetalias{darragh-ford2023} $M - c$ relation. In Fig.~\ref{fig:DeltaMass}, we present the bias in the recovered mass, expressed as $\Delta\log M = \log M_{\rm{fiducial}} - \log M_{\rm{alt}}$, as a function of the centre shift as compared to the fiducial centre, $\Delta R$. For each cluster,  measurements are presented in 10 bins in $\Delta R$, the mean value among all the random realisation in each bin is shown as the continuous line, and the standard deviation is shown as the shaded area. 
The mass bias increases with the centre offset, and reaches a plateau at around 10-15\%, for $\Delta R > 0.5 {\rm cMpc}/h$. However, we note that this plateau can be caused by the non-regularity and limited size of the BUFFALO footprints. This test would also benefit from being reproduced on a larger sample of clusters, to establish a relation between the centring error and the mass bias. Our results suggest that, in order to have an error on the mass < 2\,dex, the error on the centre position should be $< 0.2\,{\rm cMpc}/h$.

As a sanity check, we fit all the 500 miscentered realisations of the lensing signal with an off-centred halo model. As mentioned before, an off-centered term is often introduced in the models used in stacked weak lensing cluster analysis, to account for the fraction of cluster that have misidentified centres.
The off-centred term can be written as \citep[e.g][]{yang2006}:
\begin{equation}
    \Sigma_{\rm{off}}(R | M_{200}^{\rm{off}}, c^{\rm{off}}, R^{\rm{off}}) = \frac{1}{2\pi}\int_0^{2\pi} \Sigma_{\rm{NFW}}\left( \tilde{R}(\theta) | M_{200}^{\rm{off}}, c^{\rm{off}}\right)\rm{d}\theta,
    \label{equ:offcenter}
\end{equation}
where $M_{200}^{\rm{off}}$ and $c^{\rm{off}}$ are the mass and concentration of the off-centred halo, and $\tilde{R}(\theta) = \sqrt{R^2 + R_{\rm{off}}^2 - 2RR_{\rm{off}}\cos{\theta}}$, with $R_{\rm{off}}$ the projected distance between the halo centre and the centre taken to compute the lensing profile. As usual, $\bar{\Sigma}_{\rm{off}}(< R)$ is obtained by radially averaging the obtained profile. We also use the mass-concentration relation from \citetalias{darragh-ford2023} to determine the concentration from the mass, and fit two parameters, the mass, $M_{200},^{\rm off}$ and the off-centring, $R^{\rm off}$. For each cluster, we present the mass bias resulting from this off-centred modelling as dashed lines on Fig.~\ref{fig:DeltaMass}. As expected, it seems to reduce the bias, at least in the regime where the measurement is not impacted by the observation footprint edges (i.e. $\Delta R < 0.5 {\rm Mpc}/h$).

\begingroup
\renewcommand{\arraystretch}{1.5} 
\begin{table*}
    \centering
    \begin{tabular}{c | c | c c c c c c}
                            &       & A2744                  & A370                   & AS1063                 & M0416                  & M0717                  & M1149                   \\
\hline
\multirow{4}{*}{Centre}     & C0    & $14.83_{-0.06}^{+0.06}$& $15.28_{-0.04}^{+0.04}$& $14.91_{-0.05}^{+0.05}$& $14.47_{-0.06}^{+0.06}$& $15.25_{-0.05}^{+0.05}$& $14.53_{-0.08}^{+0.07}$ \\
                            & C1    & $14.83_{-0.06}^{+0.06}$& $15.30_{-0.04}^{+0.04}$& --                     & $14.53_{-0.06}^{+0.06}$& $15.30_{-0.05}^{+0.05}$& --                      \\
                            & C2    & $14.72_{-0.06}^{+0.06}$& --                     & --                     & --                     & $15.21_{-0.05}^{+0.05}$& --                      \\
                            & C3    & $14.78_{-0.06}^{+0.06}$& --                     & --                     & --                     & $14.76_{-0.09}^{+0.08}$& --                      \\
\hline 
\multirow{4}{*}{$N(z)$} & Lognorm   & $15.04_{-0.05}^{+0.05}$& $15.31_{-0.04}^{+0.04}$& $14.93_{-0.05}^{+0.05}$& $14.47_{-0.07}^{+0.06}$& $15.20_{-0.05}^{+0.05}$& $14.54_{-0.07}^{+0.07}$ \\ 
                        & S95-fit   & $15.00_{-0.05}^{+0.05}$& $15.28_{-0.04}^{+0.04}$& $14.90_{-0.05}^{+0.05}$& $14.44_{-0.06}^{+0.06}$& $15.16_{-0.05}^{+0.04}$& $14.51_{-0.07}^{+0.06}$ \\
                        & S95       & $15.35_{-0.06}^{+0.06}$& $15.66_{-0.06}^{+0.06}$& $15.23_{-0.07}^{+0.07}$& $14.71_{-0.09}^{+0.09}$& $15.68_{-0.08}^{+0.07}$& $14.95_{-0.13}^{+0.12}$ \\
                        & z mean    & $15.10_{-0.04}^{+0.04}$& $15.31_{-0.04}^{+0.03}$& $14.93_{-0.04}^{+0.04}$& $14.42_{-0.06}^{+0.05}$& $15.21_{-0.04}^{+0.04}$& $14.52_{-0.05}^{+0.05}$ \\
\hline 
\multirow{2}{*}{Fgd}    &          & $14.48_{-0.06}^{+0.06}$& $14.99_{-0.04}^{+0.04}$& $14.58_{-0.05}^{+0.05}$& $14.20_{-0.06}^{+0.06}$& $14.51_{-0.07}^{+0.06}$& $14.15_{-0.08}^{+0.07}$  \\
                        & Corrected& $15.22_{-0.05}^{+0.05}$& $15.18_{-0.05}^{+0.04}$& $14.90_{-0.05}^{+0.05}$& $14.44_{-0.07}^{+0.07}$& $15.37_{-0.05}^{+0.05}$& $14.63_{-0.08}^{+0.07}$  \\
\hline 
WL approx               &          & $15.25_{-0.06}^{+0.06}$& $15.37_{-0.05}^{+0.05}$& $14.97_{-0.06}^{+0.06}$& $14.50_{-0.07}^{+0.07}$& $15.30_{-0.05}^{+0.05}$& $14.57_{-0.09}^{+0.08}$  \\
\hline
SIS			& 	& $15.74_{-0.02}^{+0.03}$& $16.04_{-0.02}^{+0.02}$& $15.82_{-0.03}^{+0.02}$& $15.62_{-0.04}^{+0.04}$& $16.04_{-0.03}^{+0.03}$& $15.72_{-0.05}^{+0.04}$ \\
    \end{tabular}
    \caption{Summary of all masses, expressed as $\log \left(M_{200}/M_{\odot}\right)$, exploring the different measurement and modelling assumptions: the impact of the choice of the centre (Sect.~\ref{sec:miscentring}), of the background source redshift distribution $N(z)$ (Sect.~\ref{sec:Nzsyst}), of the contamination from foreground and cluster member galaxies (Sect.~\ref{sec:contaminant}), of the validity of the weak lensing approximation (Sect.~\ref{sec:wl_approx}), and of the use of a SIS density profile (Sect.~\ref{sec:SIS}).}
    \label{tab:masses_syst}
\end{table*} 
\endgroup

    \subsection{Cluster multi-modality}
    \label{sec:multimodality}
Because of their complex mass distribution, some galaxy clusters are not well described with a single halo model. The most striking case in our sample is A2744, for which the lensing profile shows a second peak, corresponding to a massive sub-structure located in the North-West of the cluster, as shown by \citet{jauzac2015b, mahler2018, Bergamini2023b, furtak2023}. Not taking this bi-modality into account when modelling the cluster weak-lensing signal may introduce a bias in the estimated cluster total mass. 

\begin{figure}
    \begin{center}
    \hspace{-5mm}\includegraphics[width=0.5\textwidth,angle=0.0]{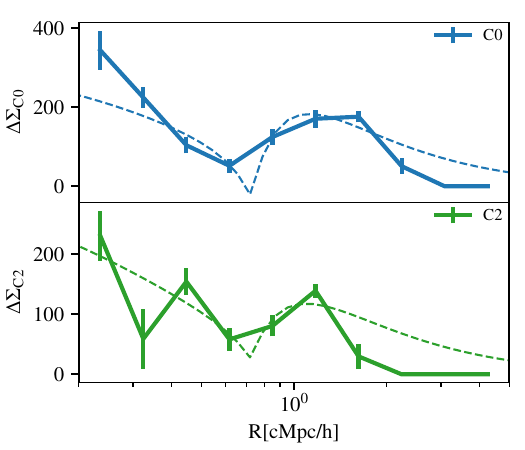}
    \caption{Excess surface mass density $\Delta\Sigma$ radial profile measured for A2744 in solid lines, centred on the BCG C0 (\emph{top panel}), and the BCG C2 (\emph{bottom panel}). The best-fit model is shown with a dashed line.}
    \label{fig:multimod}
    \end{center}
\end{figure}

To investigate this, we add a term to the mass distribution, $\Sigma_{\rm{off}}$, corresponding to a second halo, off-centred with respect to the main halo, and defined as in Equation~\ref{equ:offcenter}.
The full lensing signal is then modelled as:
\begin{multline}
    \Delta\Sigma_{\rm{tot}}(R|M_{200},M_{200}^{\rm{off}}, R^{\rm{off}},f) = \Delta\Sigma_{\rm{main}}(R | M_{200}) \\
    + f \Delta\Sigma_{\rm{off}}(R | M_{200}^{\rm{off}}, R^{\rm{off}}),
\end{multline}
and we use the mass-concentration relation from \citetalias{darragh-ford2023} to determine the concentration from the mass for both components.
In theory, the factor $f$ should be equal to 1, as both the centred and off-centred term should contribute equally to the signal\footnote{Note that it is similar but different to the miscentring term usually included in stacked weak-lensing analyses, but where the $f < 1$ factor reflects that, only for a fraction of the stacked clusters the centre is misidentified, and therefore contribute to miscentring.}. However, in our case this factor is introduced to account for the irregular footprint of the BUFFALO observations (see Fig.~\ref{fig:BUFFALO}): when measuring the shear profile in some bins at larger radial separation, missing data in some regions can prevent from azimuthally averaging over the full $360^{\circ}$, which can artificially boost the amplitude of the off-centred term.
We leave this $f$ factor as an additional free parameter. However, as it can be degenerated with the mass of the off-centred term, $M_{200}^{\rm{off}}$, we chose to optimize jointly on two shear profiles measured for the same cluster, each centred on one of the clumps in the mass distribution. 
This is shown in Fig.~\ref{fig:multimod}: the top panel shows the profile centred on the BCG C0, $\Delta\Sigma_{\rm{C0}}$ (see Fig.~\ref{fig:DSig_misc} for the position of the different BCGs), and the signal at small scales ($R \lesssim 0.5\,{\rm Mpc}/h$) corresponds to the main mass clump, while the second bump at $R \sim 1\,{\rm Mpc}/h$, to the signal of the off-centred second halo (which contains the BCG C2). Conversely, the lensing signal presented in the bottom panel, $\Delta\Sigma_{\rm{C2}}$, is centred on the BCG C2, and the second bump at $R \sim 1\,{\rm Mpc}/h$ corresponds to the halo containing C0.
The total $\chi^2$ is then $\chi^2 = \chi_1^2 +  \chi_2^2$, where
\begin{align}
    \chi_1^2 &= \frac{\left(\Delta\Sigma_{\rm{C0}}(R) - \Delta\Sigma_{\rm{tot}}(R | M_{200},M_{200}^{\rm{off}}, R^{\rm{off}},f)\right)^2}{2\sigma^2}, \\
    \chi_2^2&=\frac{\left(\Delta\Sigma_{\rm{C2}}(R) - \Delta\Sigma_{\rm{tot}}(R | M_{200}^{\rm{off}},M_{200}, R^{\rm{off}},f_2)\right)^2}{2\sigma^2},
\end{align}
and the best-fit results for the free parameters are $\log M_{200}= 14.76_{-0.07}^{+0.06}$; $\log M_{200}^{\rm{off}}=14.60_{-0.08}^{+0.07}$; $R^{\rm{off}} = 0.72_{-0.02}^{+0.02} \,{\rm cMpc}/h$; $f=3.5_{-0.5}^{+0.6}$ and $f_2=1.6_{-0.3}^{+0.3}$, and the best-fit models are shown as dashed lines on Fig.~\ref{fig:multimod}. 
We note that measuring the distance between BCGs C0 and C2 on the HST images gives $R_{\rm{C0-C2}} = 0.71\,{\rm cMpc}/h$ (in comoving units), which is in excellent agreement with the fitted value for the offset between the two components, $R^{\rm{off}} = 0.72_{-0.02}^{+0.02} \,{\rm cMpc}/h$.
Finally, if we consider that the total mass is the sum of the masses of the two sub-clusters, we obtain $\log M_{\rm{tot}} = 15.02_{-0.07}^{+0.07}$, which is, maybe surprisingly, consistent with the mass obtained when fitting a single halo model to the shear profile centred on C0. This, however, raises bigger questions, concerning the definition of a cluster. Should a (post-)merging cluster such as A2744 be considered as a single very massive cluster or as two smaller ones? This could be of importance for cluster count analyses, where consistency between the definition of a cluster in simulations and observations could potentially have an impact on the derived cosmological parameter values. 

    \subsection{Ellipticity}

As described in Sect.~\ref{sec:model_res}, the model fitted to the measured lensing signal assumes circular symmetry in the total mass distribution. In this Sect., we verify the impact of this assumption on the estimates of cluster masses, using a simplified toy model. 

\begin{figure}
    \begin{center}
    \includegraphics[width=\linewidth,angle=0.0]{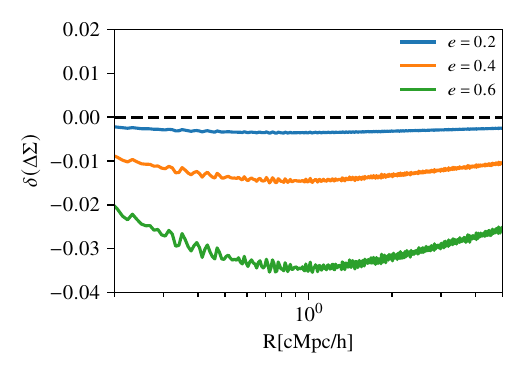}
    \caption{Relative difference on the excess surface mass density profiles $\Delta\Sigma$, between an elliptical and spherical halo of the same mass, computed as $\delta(\Delta\Sigma) = \left(\Delta\Sigma_{\rm{ellip}} - \Delta\Sigma_{\rm{circ}}\right)/\Delta\Sigma_{\rm{circ}}$, for haloes with ellipticity equal to 0.2, 0.4 and 0.6. Even for the most elliptical halo, the relative difference in the computed signals is lower than 2\%, while the statistical relative errors on the $\Delta\Sigma$ profiles measured in our sample are $\sim$ 10\%. }
    \label{fig:ellipticity}
    \end{center}
\end{figure}

We generate the projected mass map corresponding to a cluster composed of a single dark matter halo, described by a NFW density profile. We fix the mass of the halo to $M_{200} = 7 \times 10^{14}M_{\odot}$, and the concentration to $c=4$. We produce different versions of the mock cluster:  first, the fiducial version which has circular symmetry, and then we vary the ellipticity of the projected mass distribution, with $e = 0.2$, $0.4$ and $0.6$, which should represent well the distribution of cluster 2D ellipticities \citep[e.g.][]{hopkins2005}. From each mass map, we extract the radial $\Delta\Sigma$ profile, and compare the signal obtained for a circular or elliptical halo of the same mass. Figure \ref{fig:ellipticity} presents the relative differences between the measured profiles, for each of the three elliptical halos, computed as: $\delta(\Delta\Sigma) = \left(\Delta\Sigma_{\rm{ellip}} - \Delta\Sigma_{\rm{circ}}\right)/\Delta\Sigma_{\rm{circ}}$. For all the realisations of the halo, even the most elliptical one, the relative difference in the simulated lensing profile is below 4\%, which shows that assuming circular symmetry in the case of elliptical projected mass distributions has a very marginal effect on the measured signal. As a comparison, relative errors on the lensing profiles measured here for the six BUFFALO clusters are of the order of 10\%. 
Ongoing and future surveys, such as LSST and Euclid, should have lower background source densities, and therefore statistical uncertainties for individual cluster measurements no lower than those of our dataset.
On the other hand, stacked cluster analyses will have lower uncertainties in the measured shear, but stacking averages out the impact of the ellipticity of individual clusters.

We still verify the impact on the mass by fitting an NFW model to the different profiles, leaving both the mass and concentration as free parameters. We recover the input mass for all profiles, and the variation of ellipticity is only imprinted on the estimated concentration value. For the circular halo, we recover the input $c=4.0$ value, while for the elliptical haloes the measured value is slightly underestimated, with a deviation increasing with ellipticity. The effect is thus the strongest for $e=0.6$, with a concentration measured as $c=3.75$, showing that the impact of halo ellipticity is marginal.

    \subsection{Impact of $N(z)$}
    \label{sec:Nzsyst}

Another uncertainty that may impact cluster mass estimate is the imperfect knowledge of the background sources redshift distribution. As described in Sect.~\ref{sec:redshifts}, only a few percent of weakly-lensed galaxies have spectroscopically measured redshifts, around $70\%$ have less secure photometric redshifts, and the remainder have no redshift information at all. For the latter sources, we randomly draw redshifts from the measured distribution $N(z)$ when measuring the $\Delta\Sigma_r$ signal (see Sect.~\ref{sec:dsigma}). In this section, we test the impact of a possibly poorly measured $N(z)$, and of the random assignment of redshift values to sources. To be more conservative in the tests, we assume here that no individual redshifts are known for any of the sources, but only an overall $N(z)$ distribution, from which the redshifts are drawn. We consider a few different $N(z)$ distributions, and in each case follow the same process as for the fiducial analysis described in Sec.~\ref{sec:dsigma}: for each cluster and $N(z)$, we draw 100 realisations, each with a random sampling of the redshift distribution, and for each realisation further perform a bootstrap analysis to measure the statistical errors. We measure the lensing signal for each subsample source catalogue, and obtain the final measurement as the mean over all the realisations. We then fit an NFW profile, with the \citetalias{darragh-ford2023} $M-c$ relation, to obtain a $M_{200}$ estimate for each case.

We consider the following redshift distributions: (1) the fiducial lognormal distribution (see Equation~\ref{equ:lognorm}), fitted to the measured distribution of sources with known redshifts; (2) a \citet{smail1995} distribution (see Equation~\ref{equ:smail95}), with parameters fitted to the measured distribution, which does not represent well the distribution at $z < 2$; (3) a \citet{smail1995} distribution, with the parameters originally proposed in their analysis, $p_1=2$, $p_2=1$ and $p_3=5$, which represents very poorly our measured distribution (see Fig.~\ref{fig:nz}); and (4) the mean weighted redshift, $<z> = \sum_s{\left(w_s z_s\right)}/\sum_s{w_s}$, computed for each cluster individually, given in Table~\ref{tab:source_stats} and shown as dashed vertical lines in the top panel of Fig.~\ref{fig:nz}.

Resulting masses are given in Table~\ref{tab:masses_syst}, and shown in red in Fig.~\ref{fig:summary}. The different mass estimates all agree within $1\sigma$ with the fiducial measurement, except for the one performed with the ``original'' \citet{smail1995} distribution (called S95 in Table~\ref{tab:masses_syst} and Fig.~\ref{fig:summary}). This shows that (1) drawing random redshifts from a well-known $N(z)$ does not introduce significative bias in the measured mass, and (2) even an approximate knowledge of the $N(z)$, or only of the weighted mean redshift of the background galaxy sample, still leads to a consistent mass estimate. However, we note that this test does not take into account the more or less important dilution of the signal by contamination of the weak lensing catalogue by foreground or cluster member galaxies, as the background selection is performed prior to the redshift assignment. We examine this effect in the next section.


    \subsection{Dilution of the signal by foreground and cluster member galaxies}
    \label{sec:contaminant}

Ideally, the source catalogue should only contain galaxies that are located behind the galaxy cluster, as the shapes of foreground and cluster member galaxies are not affected by the gravitational potential of the lens, and therefore do not carry any weak lensing information. Including these galaxies in the catalogue will dilute the measured signal, and lead to an underestimation of the derived mass. This is particularly important as cluster member galaxies are densely populating the field, and can represent a major source of contamination. In practice, a selection is performed to remove foreground contaminants, but  usually applied selection methods, based for instance on photometric redshifts, or positions in a colour-magnitude diagram as in this analysis, are subject to different sources of uncertainty. This could lead to some residual contamination of the catalogue by foreground/cluster member galaxies.

To quantify the maximum impact that foreground contaminants can have on mass estimates, we carry out the same lensing measurement and fit procedure as described in the previous section, but using this time the full source catalogue. Specifically, we keep in the catalogue all the sources that were discarded by colour-colour and colour-magnitude cuts, and only remove bright galaxies with $m_{\rm{F814W}} < 23.5$. For all remaining galaxies, we randomly draw a redshift in the best-fit lognormal distribution, repeating this procedure 100 times for each cluster. We fit an NFW profile with the \citetalias{darragh-ford2023} $M-c$ relation to the resulting $\Delta\Sigma$ profiles; the obtained masses are given in Table~\ref{tab:masses_syst} (``Fgd''), and shown in purple in Fig.~\ref{fig:summary}. As expected, keeping foreground and cluster members in the catalogue drastically dilutes the lensing signal, leading to an underestimate of the total cluster mass.

To account for this effect, a \emph{boost factor} is often applied to the measured shear profile, correcting the dilution in the signal, such as:
\begin{equation}
    \Delta\Sigma_{\rm{corr}} (R) = \frac{\Delta\Sigma(R)}{1 - f_{\rm{cl}}(R)},
    \label{equ:boostf}
\end{equation}
where $f_{\rm cl}(R)$ is the radially dependent contamination rate of cluster/foreground galaxies in the source catalogue, and $B \equiv (1 - f_{\rm{cl}}(R))^{-1}$ is called the boost factor. Of course, the difficulty is that the contamination rate is a priori unknown, and needs to be inferred for each particular set of observations. Different strategies have been employed to characterise the contamination rate: often it is done by comparing the source density around clusters and around random points \citep[e.g.][]{sheldon2004, applegate2014, simet2017} as cluster members contaminants will cause a radially dependent overdensity. If available,  photometric redshift probability distributions of sources located near the galaxy cluster or in the field can also be used \citep[e.g.][]{melchior2017, stern2019, varga2019}.

Here, we perform a sanity check by applying a boost factor correction to the signal measured with the contaminated catalogue. As the BUFFALO observation footprints are not large enough to measure the density around random points far from the clusters, we simply use the radial dependence of the source number density. For each cluster, we then measure the source density in radial bins, using the same centre and radial spacing for the binning as for the measurement of the lensing signal. We also take into account the irregular shape of the observation footprint, as well as the masking (for instance for bright stars) to compute the densities. We then fit a decreasing exponential function, $f_{\rm dens}^{\rm exp}(R)=a*\exp{(-bR)} + c$ ($a$, $b$ and $c$ being free parameters), to the source density profile, and consider the contamination rate to simply be the source density profile normalized by its value at large radial separation, where we should be close enough to the field galaxy density, i.e. without contamination from cluster members:
\begin{equation}
    f_{\rm cl} (R) = \frac{f_{\rm dens}^{\rm exp}(R)}{f_{\rm dens}^{\rm exp}(10\,\rm Mpc)}.
\end{equation}
The corrected $\Delta\Sigma$ is then computed as in Equation~\ref{equ:boostf}, where $\Delta\Sigma$ represents the signal measured with the contaminated catalogue. Masses obtained when fitting an NFW profile to the corrected profiles are given in Table~\ref{tab:masses_syst}, and shown as the second purple point in Fig.~\ref{fig:summary} (labelled as ``Fgd-corr''). For most clusters (except A370), applying this simple boost factor allows us to correct the mass estimate within $1\sigma$ of the fiducial value, which suggests the validity of this method, as well as the purity of our fiducial catalogues.

    \subsection{Validity of the weak lensing approximation}
    \label{sec:wl_approx}

As described in Sect.~\ref{sec:model_res}, we compute the model of the lensing signal in terms of the \emph{reduced} excess surface mass density $\Delta\Sigma_r$. In some studies, the excess surface mass density can be simply considered, which assumes that the weak lensing approximation is valid, i.e that the cluster convergence is negligible, and therefore $\gamma_t \sim g_t$. This is mostly true when considering not too massive clusters, and/or when excluding the central region of the clusters from the analysis. We verify here is this approximation would be valid in our measurements. For this purpose, we fit again our fiducial weak-lensing measurement using a $\Delta\Sigma$ model \citepalias[and still using the $M-c$ relation from][]{darragh-ford2023}. The mass obtained are summarised in Table~\ref{tab:masses_syst} (`WL approx'), and shown in brown in Fig.~\ref{fig:summary}.
For some clusters (A27744, A370 and M0717), the masses are overestimated when considering the weak-lensing approximation, which is to be expected if the convergence is in fact non negligible. Indeed, for a given cluster mass, the excess surface mass density $\Delta\Sigma$ has a lower amplitude than the reduced $\Delta\Sigma_r$, leading to higher inferred masses when fitting a given measurement. This effect can also be mitigated by removing the inner cluster region, typically the central 1 Mpc, from the weak lensing analyses.

    \subsection{Density profile}
    \label{sec:SIS}

Alternatively, we use the simplest density profile to fit the lensing profiles, that is a singular isothermal sphere (SIS), for which $\Delta\Sigma$ is defined as :
\begin{equation}
    \Delta\Sigma_{\rm SIS}(R) = \frac{\sigma^2}{2 G R^2(1+z)^2},
\end{equation}
in comoving units. This profile has the advantage of having only one free parameter, the velocity dispersion $\sigma$, but is unphysical as it diverges at radius 0, and the total integrated mass diverges as well. Nevertheless, we use this simple parametrization to fit $\Delta\Sigma$ profiles, and obtain $\sigma = 1090_{-20}^{+30}$ km/s, $1534_{-32}^{+32}$ km/s, $1191_{-34}^{+32}$ km/s, $952_{-41}^{+38}$ km/s, $1537_{-48}^{+48}$ km/s, $1057_{-56}^{+45}$ km/s, for A2744, A370, AS1063, M0416, M0717 and M1149, respectively. The corresponding $\Delta\Sigma$ profiles are shown as doted-dashed green lines in Fig.~\ref{fig:DSig}.

\begin{figure*}
    \begin{center}
    \includegraphics[width=\linewidth,angle=0.0]{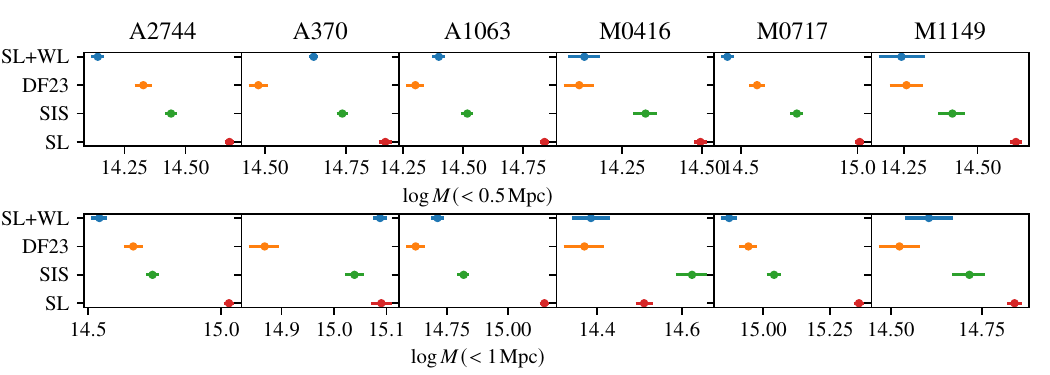}
    \caption{Mass measured within 0.5 Mpc (left panels), and within 1 Mpc (right panels), for : (1) the NFW profile fitted on the strong + weak lensing profile (blue dots) ; (2) the NFW profile with the \citetalias{darragh-ford2023} $M-c$ relation fitted on the weak lensing profile (orange dots); (3) the singular isothermal profile fitted on the weak lensing signal (green dots); and (4) the strong lensing models taken from the literature and detailed in the text.}
    \label{fig:apertures}
    \end{center}
\end{figure*}

We then compute the corresponding $M_{200}$ mass, by numerically solving $M_{\rm SIS} (<R) = \rho_{\rm crit}(z)\times \frac43 \pi R^3$. The resulting masses are given in Table~\ref{tab:masses_syst}, which shows that SIS masses are significantly larger than NFW. This is due to the fact that the SIS profile forces the density slope to be -2, which is very flat in the outskirts, therefore overestimating the amount of mass in the outer region of the cluster. To limit this effect, we make further comparison, in terms of aperture masses. We show in Fig.~\ref{fig:apertures} the masses measured within 0.5 Mpc (left panels) and 1 Mpc (right panels), and compare with values obtained when fitting the NFW profile with the \citetalias{darragh-ford2023} $M-c$ relation on the weak lensing profile (Sect.~\ref{sec:model_res}), and the NFW profile on the strong+weak lensing data (Sect.~\ref{sec:sl+wl}). For most cases, the SIS model still gives masses that are overestimated as compared to the two other models, which could suggest that this simple model lacks the flexibility to properly reproduce the observed $\Delta\Sigma$ profiles.

    \subsection{Comparison with Strong-Lensing masses}
Finally, we compare our results with masses derived from strong-lensing only models published in the literature: in Patel et al. in prep for A2744, \citet{niemiec2023} for A370, \citet{beauchesne2023} for AS1063, \citet{perera2025} for M0416, \citet{limousin2016} for M0717 and \citet{Schuldt2024} for M1149. We make the comparison in terms of aperture masses, i.e. total mass enclosed within 0.5 and 1 Mpc from the cluster centre (as defined by the X-ray emission peak). The result are presented in Fig.~\ref{fig:apertures}. 

Strong-lensing masses appear consistently larger than for other models, which could in theory be logically explained by either an over-estimation of the strong-lensing masses, or an under-estimation of the weak-lensing masses. We tend to attribute the discrepancy to the different radial ranges of validity of the models : strong-lensing constraints are typically located in the most inner few hundred kpc. Extrapolating a strong lensing model to the outer cluster regions may therefore not yield correct results. We plan for a future study on simulated clusters, to compare in details profile and mass estimates obtained from strong-lensing only, weak-lensing and combined analyses.

\begin{figure*}
    \begin{center}
    \hspace{-5mm}\includegraphics[width=\textwidth,angle=0.0]{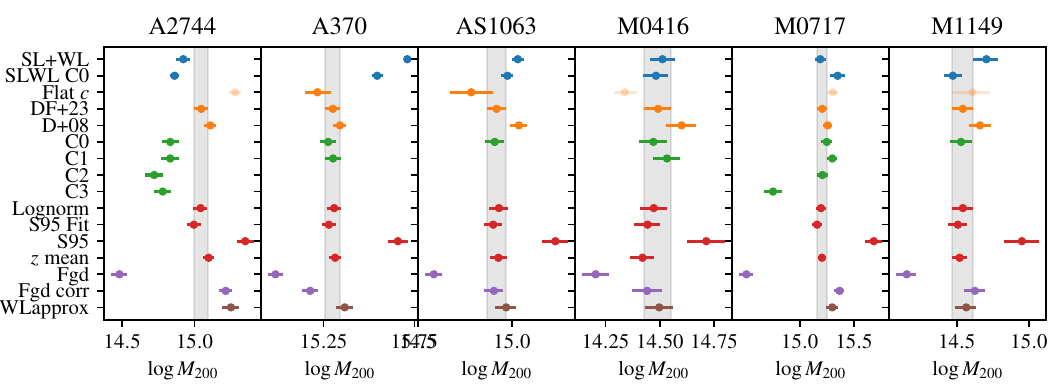}
    \caption{Summary of all masses, expressed as $\log M_{200}h/M_{\odot}$, presented in the paper, from top to bottom: strong+weak lensing measurements presented in Sect.~\ref{sec:sl+wl} (in blue, `SL+WL'); fiducial weak lensing masses measured around the X-ray peak, with different mass-concentration relations used in the NFW model as described in Sect.~\ref{sec:model_res} (in orange, `Flat $c$', `DF+23' and `D+08'); masses obtained when considering alternative centre choices as presented in Sect.~\ref{sec:miscentring} (in green, `C0', `C1', `C2' and `C3'), and different $N(z)$ distributions as described in Sect.~\ref{sec:Nzsyst} (in red, `Lognorm', `S95 Fit', `S95' and `$z$ mean'); masses measured when estimating the impact on the mass of the dilution of the weak lensing catalogue by foreground and cluster member contaminants as well as the boost-factor correction as described in Sect.~\ref{sec:contaminant} (in purple, `Fgd' for the diluted catalogue and `Fgd corr' for the boost-factor corrected signal); and the mass obtained when assuming the weak lensing approximation as presented in Sect.~\ref{sec:wl_approx}.}
    \label{fig:summary}
    \end{center}
\end{figure*}

\section{Summary and conclusions}
\label{sec:summary}

In this paper, we present weak-lensing mass measurements for six extremely massive and morphologically complex galaxy clusters, using high-resolution observations from the BUFFALO (\emph{Beyond the Ultra-deep Frontier Fields And Legacy Observations}) \emph{HST} Treasury Program. These deep space-based observations enabled the construction of high-density weak-lensing catalogues, reaching more than 50 background sources per arcmin$^2$ for each field.
We revisit and detail the methodology used to generate these weak-lensing catalogues, which are made publicly available (upon acceptance of the publication) at \url{https://archive.stsci.edu/hlsp/buffalo}. From these datasets, we extracted individual shear profiles for each cluster and derived mass estimates by fitting NFW profiles (see Table~\ref{tab:results}). Furthermore, we explored how various observational and modelling assumptions affect these mass estimates. The outcomes of these tests are summarized in Table~\ref{tab:masses_syst} and Fig.~\ref{fig:summary}, leading to the following key conclusions:
\begin{itemize}
    \item \emph{Cluster Centring:} Shear profiles were measured in radial bins centred on specific reference positions (e.g., X-ray peaks, BCG positions). We tested the sensitivity of the inferred masses to the choice of cluster centre. The results show that centring uncertainties are most problematic for strongly bimodal systems, such as A2744, which is a merger of two massive substructures (see Sect.\ref{sec:miscentring}). We also simulated random miscentring, finding, as expected, that mass uncertainties increase with the degree of offset. However, these biases can be corrected by fitting miscentred halo models (Sect.\ref{sec:random_miscentring}).
    \item \emph{Redshift Distribution Uncertainty:} Accurate measurements of the lensing profile, $\Delta\Sigma$, require knowledge of the redshift of each background source. In the absence of individual photometric redshifts, we rely on the estimated redshift distribution $N(z)$ of the source population. We randomly draw redshifts from this distribution, repeating the process 100 times to assess uncertainties. To test sensitivity to $N(z)$, we repeated the analysis with perturbed distributions and found that mass estimates are robust unless the true $N(z)$ deviates drastically from the assumed one (Sect.~\ref{sec:Nzsyst}).
    \item \emph{Foreground Contamination and Boost Factor Correction:} Only background galaxies contribute to the lensing signal. Without proper background selection, the signal is diluted, leading to underestimated masses. We demonstrate that this effect is generally well-corrected by applying a boost-factor method, which accounts for foreground contamination statistically (Sect.~\ref{sec:contaminant}).
    \item \emph{Validity of the Weak-Lensing Approximation}: Given the high masses of these clusters, magnification effects near the core regions can violate the weak-lensing assumption, where the reduced tangential shear is close to the true shear $g_{\rm t}\approx \gamma_{\rm t}$. Although we exclude the most highly magnified inner regions based on an empirical cut, we examined whether residual biases might remain. We tested two alternative correction strategies, but the results remain inconclusive; the possibility of a residual bias should be kept in mind (Sect.~\ref{sec:wl_approx}).
\end{itemize}

In conclusion, we assessed the influence of several observational and modeling effects on weak-lensing mass estimates. We find that the largest sources of uncertainty arise for disturbed and merging systems such as A2744. This raises fundamental questions: should such systems be treated as a single cluster or multiple halos? What is their prevalence as a function of mass and redshift? These are crucial questions for future work, both observational- and simulation-based, especially given the impact that such systems may have on cosmological analyses.

\section*{Acknowledgments}
AA acknowledges financial support through the Beatriz Galindo programme and the project PID2022-138896NB-C51 (MCIU/AEI/MINECO/FEDER, UE), Ministerio de Ciencia, Investigación y Universidades. MJ acknowledges support by the United Kingdom Research and Innovation (UKRI) Future Leaders Fellowship `Using Cosmic Beasts to uncover the Nature of Dark Matter' (grant number MR/X006069/1).
BB acknowledges the Swiss National Science Foundation (SNSF) for support.
JMD acknowledges the support of projects PID2022-138896NB-C51 (MCIU/AEI/MINECO/FEDER, UE) Ministerio de Ciencia, Investigaci\'on y Universidades and SA101P24.
ML acknowledges the Centre National de la Recherche Scientifique (CNRS) and the Centre National des Etudes Spatiale (CNES) for support.


\bibliographystyle{mnras}
\bibliography{lensing.bib}

\begin{thebibliography}{}
\makeatletter
\relax
\def\mn@urlcharsother{\let\do\@makeother \do\$\do\&\do\#\do\^\do\_\do\%\do\~}
\def\mn@doi{\begingroup\mn@urlcharsother \@ifnextchar [ {\mn@doi@}
  {\mn@doi@[]}}
\def\mn@doi@[#1]#2{\def\@tempa{#1}\ifx\@tempa\@empty \href
  {http://dx.doi.org/#2} {doi:#2}\else \href {http://dx.doi.org/#2} {#1}\fi
  \endgroup}
\def\mn@eprint#1#2{\mn@eprint@#1:#2::\@nil}
\def\mn@eprint@arXiv#1{\href {http://arxiv.org/abs/#1} {{\tt arXiv:#1}}}
\def\mn@eprint@dblp#1{\href {http://dblp.uni-trier.de/rec/bibtex/#1.xml}
  {dblp:#1}}
\def\mn@eprint@#1:#2:#3:#4\@nil{\def\@tempa {#1}\def\@tempb {#2}\def\@tempc
  {#3}\ifx \@tempc \@empty \let \@tempc \@tempb \let \@tempb \@tempa \fi \ifx
  \@tempb \@empty \def\@tempb {arXiv}\fi \@ifundefined
  {mn@eprint@\@tempb}{\@tempb:\@tempc}{\expandafter \expandafter \csname
  mn@eprint@\@tempb\endcsname \expandafter{\@tempc}}}

\bibitem[\protect\citeauthoryear{{Abriola}, {Della Pergola}, {Lombardi},
  {Bergamini}, {Nonino}, {Grillo}  \& {Rosati}}{{Abriola}
  et~al.}{2024}]{Abriola2024}
{Abriola} D.,  {Della Pergola} D.,  {Lombardi} M.,  {Bergamini} P.,  {Nonino}
  M.,  {Grillo} C.,   {Rosati} P.,  2024, \mn@doi [\aap]
  {10.1051/0004-6361/202347887}, \href
  {https://ui.adsabs.harvard.edu/abs/2024A&A...684A.193A} {684, A193}

\bibitem[\protect\citeauthoryear{{Applegate} et~al.,}{{Applegate}
  et~al.}{2014}]{applegate2014}
{Applegate} D.~E.,  et~al., 2014, \mn@doi [\mnras] {10.1093/mnras/stt2129},
  \href {https://ui.adsabs.harvard.edu/abs/2014MNRAS.439...48A} {439, 48}

\bibitem[\protect\citeauthoryear{{Arnaud}}{{Arnaud}}{1996}]{arnaud1996}
{Arnaud} K.~A.,  1996, in {Jacoby} G.~H.,  {Barnes} J.,  eds,  Astronomical
  Society of the Pacific Conference Series Vol. 101, Astronomical Data Analysis
  Software and Systems V. p.~17

\bibitem[\protect\citeauthoryear{{Arnouts}, {Cristiani}, {Moscardini},
  {Matarrese}, {Lucchin}, {Fontana}  \& {Giallongo}}{{Arnouts}
  et~al.}{1999}]{arnouts1999}
{Arnouts} S.,  {Cristiani} S.,  {Moscardini} L.,  {Matarrese} S.,  {Lucchin}
  F.,  {Fontana} A.,   {Giallongo} E.,  1999, \mn@doi [\mnras]
  {10.1046/j.1365-8711.1999.02978.x}, \href
  {http://adsabs.harvard.edu/abs/1999MNRAS.310..540A} {310, 540}

\bibitem[\protect\citeauthoryear{{Beauchesne} et~al.,}{{Beauchesne}
  et~al.}{2024}]{beauchesne2023}
{Beauchesne} B.,  et~al., 2024, \mn@doi [\mnras] {10.1093/mnras/stad3308},
  \href {https://ui.adsabs.harvard.edu/abs/2024MNRAS.527.3246B} {527, 3246}

\bibitem[\protect\citeauthoryear{{Bergamini} et~al.,}{{Bergamini}
  et~al.}{2023a}]{bergamini2023}
{Bergamini} P.,  et~al., 2023a, \mn@doi [\aap] {10.1051/0004-6361/202244834},
  \href {https://ui.adsabs.harvard.edu/abs/2023A&A...674A..79B} {674, A79}

\bibitem[\protect\citeauthoryear{{Bergamini} et~al.,}{{Bergamini}
  et~al.}{2023b}]{Bergamini2023b}
{Bergamini} P.,  et~al., 2023b, \mn@doi [\apj] {10.3847/1538-4357/acd643},
  \href {https://ui.adsabs.harvard.edu/abs/2023ApJ...952...84B} {952, 84}

\bibitem[\protect\citeauthoryear{{Bertin} \& {Arnouts}}{{Bertin} \&
  {Arnouts}}{1996}]{bertin1996}
{Bertin} E.,  {Arnouts} S.,  1996, \mn@doi [\aaps] {10.1051/aas:1996164}, \href
  {https://ui.adsabs.harvard.edu/abs/1996A&AS..117..393B} {117, 393}

\bibitem[\protect\citeauthoryear{{Darragh-Ford}, {Mantz}, {Rasia}, {Allen},
  {Morris}, {Foster}, {Schmidt}  \& {Wenrich}}{{Darragh-Ford}
  et~al.}{2023}]{darragh-ford2023}
{Darragh-Ford} E.,  {Mantz} A.~B.,  {Rasia} E.,  {Allen} S.~W.,  {Morris}
  R.~G.,  {Foster} J.,  {Schmidt} R.~W.,   {Wenrich} G.,  2023, \mn@doi
  [\mnras] {10.1093/mnras/stad585}, \href
  {https://ui.adsabs.harvard.edu/abs/2023MNRAS.521..790D} {521, 790}

\bibitem[\protect\citeauthoryear{{Diego}, {Protopapas}, {Sandvik}  \&
  {Tegmark}}{{Diego} et~al.}{2005}]{diego2005}
{Diego} J.~M.,  {Protopapas} P.,  {Sandvik} H.~B.,   {Tegmark} M.,  2005,
  \mn@doi [Monthly Notices of the Royal Astronomical Society]
  {10.1111/j.1365-2966.2005.09021.x}, \href
  {https://ui.adsabs.harvard.edu/abs/2005MNRAS.360..477D} {360, 477}

\bibitem[\protect\citeauthoryear{{Diego} et~al.,}{{Diego}
  et~al.}{2016}]{diego2016}
{Diego} J.~M.,  et~al., 2016, \mn@doi [Monthly Notices of the Royal
  Astronomical Society] {10.1093/mnras/stv2638}, \href
  {https://ui.adsabs.harvard.edu/abs/2016MNRAS.456..356D} {456, 356}

\bibitem[\protect\citeauthoryear{{Diego} et~al.,}{{Diego}
  et~al.}{2024}]{Diego2024}
{Diego} J.~M.,  et~al., 2024, \mn@doi [\aap] {10.1051/0004-6361/202349119},
  \href {https://ui.adsabs.harvard.edu/abs/2024A&A...690A.114D} {690, A114}

\bibitem[\protect\citeauthoryear{{Duffy}, {Schaye}, {Kay}  \& {Dalla
  Vecchia}}{{Duffy} et~al.}{2008}]{duffy2008}
{Duffy} A.~R.,  {Schaye} J.,  {Kay} S.~T.,   {Dalla Vecchia} C.,  2008, \mn@doi
  [\mnras] {10.1111/j.1745-3933.2008.00537.x}, \href
  {http://adsabs.harvard.edu/abs/2008MNRAS.390L..64D} {390, L64}

\bibitem[\protect\citeauthoryear{{Eckert}, {Ettori}, {Pointecouteau},
  {Molendi}, {Paltani}  \& {Tchernin}}{{Eckert} et~al.}{2017}]{xcop}
{Eckert} D.,  {Ettori} S.,  {Pointecouteau} E.,  {Molendi} S.,  {Paltani} S.,
  {Tchernin} C.,  2017, \mn@doi [Astronomische Nachrichten]
  {10.1002/asna.201713345}, \href
  {https://ui.adsabs.harvard.edu/abs/2017AN....338..293E} {338, 293}

\bibitem[\protect\citeauthoryear{{Eckert}, {Finoguenov}, {Ghirardini},
  {Grandis}, {Kaefer}, {Sanders}  \& {Ramos-Ceja}}{{Eckert}
  et~al.}{2020}]{eckert2020}
{Eckert} D.,  {Finoguenov} A.,  {Ghirardini} V.,  {Grandis} S.,  {Kaefer} F.,
  {Sanders} J.,   {Ramos-Ceja} M.,  2020, \mn@doi [The Open Journal of
  Astrophysics] {10.21105/astro.2009.13944}, \href
  {https://ui.adsabs.harvard.edu/abs/2020OJAp....3E..12E} {3, 12}

\bibitem[\protect\citeauthoryear{{Euclid Collaboration} et~al.,}{{Euclid
  Collaboration} et~al.}{2022}]{euclid2022}
{Euclid Collaboration} et~al., 2022, \mn@doi [\aap]
  {10.1051/0004-6361/202141938}, \href
  {https://ui.adsabs.harvard.edu/abs/2022A&A...662A.112E} {662, A112}

\bibitem[\protect\citeauthoryear{{Euclid Collaboration} et~al.,}{{Euclid
  Collaboration} et~al.}{2025}]{mellier2025}
{Euclid Collaboration} et~al., 2025, \mn@doi [\aap]
  {10.1051/0004-6361/202450810}, \href
  {https://ui.adsabs.harvard.edu/abs/2025A&A...697A...1E} {697, A1}

\bibitem[\protect\citeauthoryear{{Foreman-Mackey}, {Hogg}, {Lang}  \&
  {Goodman}}{{Foreman-Mackey} et~al.}{2013}]{foreman-mackey2013}
{Foreman-Mackey} D.,  {Hogg} D.~W.,  {Lang} D.,   {Goodman} J.,  2013, \mn@doi
  [\pasp] {10.1086/670067}, \href
  {http://adsabs.harvard.edu/abs/2013PASP..125..306F} {125, 306}

\bibitem[\protect\citeauthoryear{{Fruscione} et~al.,}{{Fruscione}
  et~al.}{2006}]{ciao2006}
{Fruscione} A.,  et~al., 2006, in {Silva} D.~R.,  {Doxsey} R.~E.,  eds,
  Society of Photo-Optical Instrumentation Engineers (SPIE) Conference Series
  Vol. 6270, Society of Photo-Optical Instrumentation Engineers (SPIE)
  Conference Series. p. 62701V, \mn@doi{10.1117/12.671760}

\bibitem[\protect\citeauthoryear{{Furtak} et~al.,}{{Furtak}
  et~al.}{2023}]{furtak2023}
{Furtak} L.~J.,  et~al., 2023, \mn@doi [\mnras] {10.1093/mnras/stad1627}, \href
  {https://ui.adsabs.harvard.edu/abs/2023MNRAS.523.4568F} {523, 4568}

\bibitem[\protect\citeauthoryear{{Ghirardini}, {Ettori}, {Eckert}  \&
  {Molendi}}{{Ghirardini} et~al.}{2019}]{ghirardini2019}
{Ghirardini} V.,  {Ettori} S.,  {Eckert} D.,   {Molendi} S.,  2019, \mn@doi
  [\aap] {10.1051/0004-6361/201834875}, \href
  {https://ui.adsabs.harvard.edu/abs/2019A&A...627A..19G} {627, A19}

\bibitem[\protect\citeauthoryear{{Giocoli}, {Meneghetti}, {Bartelmann},
  {Moscardini}  \& {Boldrin}}{{Giocoli} et~al.}{2012}]{giocoli2012a}
{Giocoli} C.,  {Meneghetti} M.,  {Bartelmann} M.,  {Moscardini} L.,   {Boldrin}
  M.,  2012, \mn@doi [\mnras] {10.1111/j.1365-2966.2012.20558.x}, \href
  {https://ui.adsabs.harvard.edu/abs/2012MNRAS.421.3343G} {421, 3343}

\bibitem[\protect\citeauthoryear{Gladders \& Yee}{Gladders \&
  Yee}{2000}]{gladders2000}
Gladders M.~D.,  Yee H. K.~C.,  2000, \mn@doi [The Astronomical Journal]
  {10.1086/301557}, 120, 2148

\bibitem[\protect\citeauthoryear{{Grillo} et~al.,}{{Grillo}
  et~al.}{2016}]{Grillo2016}
{Grillo} C.,  et~al., 2016, \mn@doi [\apj] {10.3847/0004-637X/822/2/78}, \href
  {https://ui.adsabs.harvard.edu/abs/2016ApJ...822...78G} {822, 78}

\bibitem[\protect\citeauthoryear{{Grogin} et~al.,}{{Grogin}
  et~al.}{2011}]{grogin2011}
{Grogin} N.~A.,  et~al., 2011, \mn@doi [\apjs] {10.1088/0067-0049/197/2/35},
  \href {https://ui.adsabs.harvard.edu/abs/2011ApJS..197...35G} {197, 35}

\bibitem[\protect\citeauthoryear{{Harvey} \& {Massey}}{{Harvey} \&
  {Massey}}{2024}]{Harvey2024}
{Harvey} D.~R.,  {Massey} R.,  2024, \mn@doi [\mnras] {10.1093/mnras/stae370},
  \href {https://ui.adsabs.harvard.edu/abs/2024MNRAS.529..802H} {529, 802}

\bibitem[\protect\citeauthoryear{{Harvey}, {Tam}, {Jauzac}, {Massey}  \&
  {Rhodes}}{{Harvey} et~al.}{2019}]{harvey2019}
{Harvey} D.,  {Tam} S.-I.,  {Jauzac} M.,  {Massey} R.,   {Rhodes} J.,  2019,
  arXiv e-prints, \href {https://ui.adsabs.harvard.edu/abs/2019arXiv191106333H}
  {p. arXiv:1911.06333}

\bibitem[\protect\citeauthoryear{Hopkins, Bahcall  \& Bode}{Hopkins
  et~al.}{2005}]{hopkins2005}
Hopkins P.~F.,  Bahcall N.~A.,   Bode P.,  2005, \mn@doi [The Astrophysical
  Journal] {10.1086/425993}, 618, 1

\bibitem[\protect\citeauthoryear{{Ilbert} et~al.,}{{Ilbert}
  et~al.}{2006}]{ilbert2006}
{Ilbert} O.,  et~al., 2006, \mn@doi [\aap] {10.1051/0004-6361:20065138}, \href
  {http://adsabs.harvard.edu/abs/2006A%26A...457..841I} {457, 841}

\bibitem[\protect\citeauthoryear{{Ivezi{\'c}} et~al.,}{{Ivezi{\'c}}
  et~al.}{2019}]{ivezic2019}
{Ivezi{\'c}} {\v{Z}}.,  et~al., 2019, \mn@doi [\apj]
  {10.3847/1538-4357/ab042c}, \href
  {https://ui.adsabs.harvard.edu/abs/2019ApJ...873..111I} {873, 111}

\bibitem[\protect\citeauthoryear{{Jauzac} et~al.,}{{Jauzac}
  et~al.}{2012}]{jauzac2012}
{Jauzac} M.,  et~al., 2012, \mn@doi [\mnras]
  {10.1111/j.1365-2966.2012.21966.x}, \href
  {https://ui.adsabs.harvard.edu/abs/2012MNRAS.426.3369J} {426, 3369}

\bibitem[\protect\citeauthoryear{{Jauzac} et~al.,}{{Jauzac}
  et~al.}{2015a}]{jauzac2015a}
{Jauzac} M.,  et~al., 2015a, \mn@doi [\mnras] {10.1093/mnras/stu2425}, \href
  {https://ui.adsabs.harvard.edu/abs/2015MNRAS.446.4132J} {446, 4132}

\bibitem[\protect\citeauthoryear{{Jauzac} et~al.,}{{Jauzac}
  et~al.}{2015b}]{jauzac2015b}
{Jauzac} M.,  et~al., 2015b, \mn@doi [\mnras] {10.1093/mnras/stv1402}, \href
  {https://ui.adsabs.harvard.edu/abs/2015MNRAS.452.1437J} {452, 1437}

\bibitem[\protect\citeauthoryear{{Jauzac} et~al.,}{{Jauzac}
  et~al.}{2016a}]{jauzac2016a}
{Jauzac} M.,  et~al., 2016a, \mn@doi [\mnras] {10.1093/mnras/stw069}, \href
  {https://ui.adsabs.harvard.edu/abs/2016MNRAS.457.2029J} {457, 2029}

\bibitem[\protect\citeauthoryear{{Jauzac} et~al.,}{{Jauzac}
  et~al.}{2016b}]{jauzac2016c}
{Jauzac} M.,  et~al., 2016b, \mn@doi [\mnras] {10.1093/mnras/stw2251}, \href
  {https://ui.adsabs.harvard.edu/abs/2016MNRAS.463.3876J} {463, 3876}

\bibitem[\protect\citeauthoryear{{Jullo}, {Kneib}, {Limousin},
  {El{\'{\i}}asd{\'o}ttir}, {Marshall}  \& {Verdugo}}{{Jullo}
  et~al.}{2007}]{jullo2007}
{Jullo} E.,  {Kneib} J.-P.,  {Limousin} M.,  {El{\'{\i}}asd{\'o}ttir} {\'A}.,
  {Marshall} P.~J.,   {Verdugo} T.,  2007, \mn@doi [New Journal of Physics]
  {10.1088/1367-2630/9/12/447}, \href
  {http://adsabs.harvard.edu/abs/2007NJPh....9..447J} {9, 447}

\bibitem[\protect\citeauthoryear{{Katgert}, {Biviano}  \& {Mazure}}{{Katgert}
  et~al.}{2004}]{katgert2004}
{Katgert} P.,  {Biviano} A.,   {Mazure} A.,  2004, \mn@doi [\apj]
  {10.1086/380118}, \href
  {https://ui.adsabs.harvard.edu/abs/2004ApJ...600..657K} {600, 657}

\bibitem[\protect\citeauthoryear{{Koekemoer} et~al.,}{{Koekemoer}
  et~al.}{2011}]{koekemoer2011}
{Koekemoer} A.~M.,  et~al., 2011, \mn@doi [\apjs] {10.1088/0067-0049/197/2/36},
  \href {https://ui.adsabs.harvard.edu/abs/2011ApJS..197...36K} {197, 36}

\bibitem[\protect\citeauthoryear{{Krist}, {Hook}  \& {Stoehr}}{{Krist}
  et~al.}{2011}]{krist2011}
{Krist} J.~E.,  {Hook} R.~N.,   {Stoehr} F.,  2011, in {Kahan} M.~A.,  ed.,
  Society of Photo-Optical Instrumentation Engineers (SPIE) Conference Series
  Vol. 8127, Optical Modeling and Performance Predictions V. p. 81270J,
  \mn@doi{10.1117/12.892762}

\bibitem[\protect\citeauthoryear{{Lagattuta} et~al.,}{{Lagattuta}
  et~al.}{2019}]{lagattuta2019}
{Lagattuta} D.~J.,  et~al., 2019, \mn@doi [Monthly Notices of the Royal
  Astronomical Society] {10.1093/mnras/stz620}, \href
  {https://ui.adsabs.harvard.edu/abs/2019MNRAS.485.3738L} {485, 3738}

\bibitem[\protect\citeauthoryear{{Lagattuta} et~al.,}{{Lagattuta}
  et~al.}{2022}]{lagattuta2022}
{Lagattuta} D.~J.,  et~al., 2022, \mn@doi [\mnras] {10.1093/mnras/stac418},
  \href {https://ui.adsabs.harvard.edu/abs/2022MNRAS.514..497L} {514, 497}

\bibitem[\protect\citeauthoryear{{Laureijs} et~al.,}{{Laureijs}
  et~al.}{2011}]{laureijs2011}
{Laureijs} R.,  et~al., 2011, \mn@doi [arXiv e-prints]
  {10.48550/arXiv.1110.3193}, \href
  {https://ui.adsabs.harvard.edu/abs/2011arXiv1110.3193L} {p. arXiv:1110.3193}

\bibitem[\protect\citeauthoryear{{Leauthaud} et~al.,}{{Leauthaud}
  et~al.}{2007}]{leauthaud2007}
{Leauthaud} A.,  et~al., 2007, \mn@doi [\apjs] {10.1086/516598}, \href
  {http://adsabs.harvard.edu/abs/2007ApJS..172..219L} {172, 219}

\bibitem[\protect\citeauthoryear{{Limousin} et~al.,}{{Limousin}
  et~al.}{2016}]{limousin2016}
{Limousin} M.,  et~al., 2016, \mn@doi [\aap] {10.1051/0004-6361/201527638},
  \href {http://adsabs.harvard.edu/abs/2016A%26A...588A..99L} {588, A99}

\bibitem[\protect\citeauthoryear{{Lotz} et~al.,}{{Lotz}
  et~al.}{2017}]{lotz2017}
{Lotz} J.~M.,  et~al., 2017, \mn@doi [\apj] {10.3847/1538-4357/837/1/97}, \href
  {https://ui.adsabs.harvard.edu/abs/2017ApJ...837...97L} {837, 97}

\bibitem[\protect\citeauthoryear{{Mahler} et~al.,}{{Mahler}
  et~al.}{2018}]{mahler2018}
{Mahler} G.,  et~al., 2018, \mn@doi [\mnras] {10.1093/mnras/stx1971}, \href
  {https://ui.adsabs.harvard.edu/abs/2018MNRAS.473..663M} {473, 663}

\bibitem[\protect\citeauthoryear{{Medezinski}, {Umetsu}, {Okabe}, {Nonino},
  {Molnar}, {Massey}, {Dupke}  \& {Merten}}{{Medezinski}
  et~al.}{2016}]{Medezinski2016}
{Medezinski} E.,  {Umetsu} K.,  {Okabe} N.,  {Nonino} M.,  {Molnar} S.,
  {Massey} R.,  {Dupke} R.,   {Merten} J.,  2016, \mn@doi [\apj]
  {10.3847/0004-637X/817/1/24}, \href
  {https://ui.adsabs.harvard.edu/abs/2016ApJ...817...24M} {817, 24}

\bibitem[\protect\citeauthoryear{{Melchior} et~al.,}{{Melchior}
  et~al.}{2017}]{melchior2017}
{Melchior} P.,  et~al., 2017, \mn@doi [\mnras] {10.1093/mnras/stx1053}, \href
  {https://ui.adsabs.harvard.edu/abs/2017MNRAS.469.4899M} {469, 4899}

\bibitem[\protect\citeauthoryear{{Merten} et~al.,}{{Merten}
  et~al.}{2015}]{merten2015}
{Merten} J.,  et~al., 2015, \mn@doi [\apj] {10.1088/0004-637X/806/1/4}, \href
  {https://ui.adsabs.harvard.edu/abs/2015ApJ...806....4M} {806, 4}

\bibitem[\protect\citeauthoryear{{Navarro}, {Frenk}  \& {White}}{{Navarro}
  et~al.}{1996}]{nfw1996}
{Navarro} J.~F.,  {Frenk} C.~S.,   {White} S.~D.~M.,  1996, \mn@doi [\apj]
  {10.1086/177173}, \href {http://adsabs.harvard.edu/abs/1996ApJ...462..563N}
  {462, 563}

\bibitem[\protect\citeauthoryear{{Niemiec} et~al.,}{{Niemiec}
  et~al.}{2023}]{niemiec2023}
{Niemiec} A.,  et~al., 2023, \mn@doi [\mnras] {10.1093/mnras/stad1999}, \href
  {https://ui.adsabs.harvard.edu/abs/2023MNRAS.524.2883N} {524, 2883}

\bibitem[\protect\citeauthoryear{{Oguri}}{{Oguri}}{2010}]{oguri2010}
{Oguri} M.,  2010, \mn@doi [\pasj] {10.1093/pasj/62.4.1017}, \href
  {https://ui.adsabs.harvard.edu/abs/2010PASJ...62.1017O} {62, 1017}

\bibitem[\protect\citeauthoryear{{Pagul} et~al.,}{{Pagul}
  et~al.}{2024}]{pagul2024}
{Pagul} A.,  et~al., 2024, \mn@doi [\apjs] {10.3847/1538-4365/ad40a1}, \href
  {https://ui.adsabs.harvard.edu/abs/2024ApJS..273...10P} {273, 10}

\bibitem[\protect\citeauthoryear{{Perera} et~al.,}{{Perera}
  et~al.}{2025}]{perera2025}
{Perera} D.,  et~al., 2025, \mn@doi [\mnras] {10.1093/mnras/stae2753}, \href
  {https://ui.adsabs.harvard.edu/abs/2025MNRAS.536.2690P} {536, 2690}

\bibitem[\protect\citeauthoryear{{Planck Collaboration} et~al.,}{{Planck
  Collaboration} et~al.}{2020}]{planck2018}
{Planck Collaboration} et~al., 2020, \mn@doi [\aap]
  {10.1051/0004-6361/201833910}, \href
  {https://ui.adsabs.harvard.edu/abs/2020A&A...641A...6P} {641, A6}

\bibitem[\protect\citeauthoryear{{Pratt} \& {Arnaud}}{{Pratt} \&
  {Arnaud}}{2002}]{pratt2002}
{Pratt} G.~W.,  {Arnaud} M.,  2002, \mn@doi [\aap]
  {10.1051/0004-6361:20021032}, \href
  {https://ui.adsabs.harvard.edu/abs/2002A&A...394..375P} {394, 375}

\bibitem[\protect\citeauthoryear{{Pratt}, {Arnaud}, {Biviano}, {Eckert},
  {Ettori}, {Nagai}, {Okabe}  \& {Reiprich}}{{Pratt} et~al.}{2019}]{pratt2019}
{Pratt} G.~W.,  {Arnaud} M.,  {Biviano} A.,  {Eckert} D.,  {Ettori} S.,
  {Nagai} D.,  {Okabe} N.,   {Reiprich} T.~H.,  2019, \mn@doi [\ssr]
  {10.1007/s11214-019-0591-0}, \href
  {https://ui.adsabs.harvard.edu/abs/2019SSRv..215...25P} {215, 25}

\bibitem[\protect\citeauthoryear{{Rhodes}, {Refregier}  \& {Groth}}{{Rhodes}
  et~al.}{2000}]{rhodes2000}
{Rhodes} J.,  {Refregier} A.,   {Groth} E.~J.,  2000, \mn@doi [\apj]
  {10.1086/308902}, \href
  {https://ui.adsabs.harvard.edu/abs/2000ApJ...536...79R} {536, 79}

\bibitem[\protect\citeauthoryear{{Richard} et~al.,}{{Richard}
  et~al.}{2021}]{Richard2021}
{Richard} J.,  et~al., 2021, \mn@doi [\aap] {10.1051/0004-6361/202039462},
  \href {https://ui.adsabs.harvard.edu/abs/2021A&A...646A..83R} {646, A83}

\bibitem[\protect\citeauthoryear{{Rihtar{\v{s}}i{\v{c}}}
  et~al.,}{{Rihtar{\v{s}}i{\v{c}}} et~al.}{2025}]{Rihtarsic2025}
{Rihtar{\v{s}}i{\v{c}}} G.,  et~al., 2025, \mn@doi [\aap]
  {10.1051/0004-6361/202451117}, \href
  {https://ui.adsabs.harvard.edu/abs/2025A&A...696A..15R} {696, A15}

\bibitem[\protect\citeauthoryear{{Rossetti} et~al.,}{{Rossetti}
  et~al.}{2024}]{rossetti2024}
{Rossetti} M.,  et~al., 2024, \mn@doi [\aap] {10.1051/0004-6361/202348853},
  \href {https://ui.adsabs.harvard.edu/abs/2024A&A...686A..68R} {686, A68}

\bibitem[\protect\citeauthoryear{{Sarazin}}{{Sarazin}}{1988}]{sarazin1988}
{Sarazin} C.~L.,  1988, {X-ray emission from clusters of galaxies}

\bibitem[\protect\citeauthoryear{{Schmidt} et~al.,}{{Schmidt}
  et~al.}{2014}]{Schmidt2014}
{Schmidt} K.~B.,  et~al., 2014, \mn@doi [\apjl] {10.1088/2041-8205/782/2/L36},
  \href {https://ui.adsabs.harvard.edu/abs/2014ApJ...782L..36S} {782, L36}

\bibitem[\protect\citeauthoryear{{Schuldt} et~al.,}{{Schuldt}
  et~al.}{2024}]{Schuldt2024}
{Schuldt} S.,  et~al., 2024, \mn@doi [\aap] {10.1051/0004-6361/202449528},
  \href {https://ui.adsabs.harvard.edu/abs/2024A&A...689A..42S} {689, A42}

\bibitem[\protect\citeauthoryear{{Sereno}}{{Sereno}}{2025}]{sereno2025}
{Sereno} M.,  2025, \mn@doi [\aap] {10.1051/0004-6361/202553989}, \href
  {https://ui.adsabs.harvard.edu/abs/2025A&A...696A.227S} {696, A227}

\bibitem[\protect\citeauthoryear{{Sheldon} et~al.,}{{Sheldon}
  et~al.}{2004}]{sheldon2004}
{Sheldon} E.~S.,  et~al., 2004, \mn@doi [\aj] {10.1086/383293}, \href
  {https://ui.adsabs.harvard.edu/abs/2004AJ....127.2544S} {127, 2544}

\bibitem[\protect\citeauthoryear{{Simet}, {McClintock}, {Mandelbaum}, {Rozo},
  {Rykoff}, {Sheldon}  \& {Wechsler}}{{Simet} et~al.}{2017}]{simet2017}
{Simet} M.,  {McClintock} T.,  {Mandelbaum} R.,  {Rozo} E.,  {Rykoff} E.,
  {Sheldon} E.,   {Wechsler} R.~H.,  2017, \mn@doi [\mnras]
  {10.1093/mnras/stw3250}, \href
  {http://adsabs.harvard.edu/abs/2017MNRAS.466.3103S} {466, 3103}

\bibitem[\protect\citeauthoryear{{Smail} \& {Dickinson}}{{Smail} \&
  {Dickinson}}{1995}]{smail1995}
{Smail} I.,  {Dickinson} M.,  1995, \mn@doi [\apjl] {10.1086/309842}, \href
  {https://ui.adsabs.harvard.edu/abs/1995ApJ...455L..99S} {455, L99}

\bibitem[\protect\citeauthoryear{{Steinhardt} et~al.,}{{Steinhardt}
  et~al.}{2020}]{steinhardt2020}
{Steinhardt} C.~L.,  et~al., 2020, \mn@doi [\apjs] {10.3847/1538-4365/ab75ed},
  \href {https://ui.adsabs.harvard.edu/abs/2020ApJS..247...64S} {247, 64}

\bibitem[\protect\citeauthoryear{{Stern} et~al.,}{{Stern}
  et~al.}{2019}]{stern2019}
{Stern} C.,  et~al., 2019, \mn@doi [\mnras] {10.1093/mnras/stz234}, \href
  {https://ui.adsabs.harvard.edu/abs/2019MNRAS.485...69S} {485, 69}

\bibitem[\protect\citeauthoryear{{Sunyaev} \& {Zeldovich}}{{Sunyaev} \&
  {Zeldovich}}{1972}]{sunyaev1972}
{Sunyaev} R.~A.,  {Zeldovich} Y.~B.,  1972, Comments on Astrophysics and Space
  Physics, \href {https://ui.adsabs.harvard.edu/abs/1972CoASP...4..173S} {4,
  173}

\bibitem[\protect\citeauthoryear{{Suyu}, {Marshall}, {Auger}, {Hilbert},
  {Blandford}, {Koopmans}, {Fassnacht}  \& {Treu}}{{Suyu}
  et~al.}{2010}]{Suyu2010}
{Suyu} S.~H.,  {Marshall} P.~J.,  {Auger} M.~W.,  {Hilbert} S.,  {Blandford}
  R.~D.,  {Koopmans} L.~V.~E.,  {Fassnacht} C.~D.,   {Treu} T.,  2010, \mn@doi
  [\apj] {10.1088/0004-637X/711/1/201}, \href
  {https://ui.adsabs.harvard.edu/abs/2010ApJ...711..201S} {711, 201}

\bibitem[\protect\citeauthoryear{{Suyu} et~al.,}{{Suyu}
  et~al.}{2012}]{Suyu2012}
{Suyu} S.~H.,  et~al., 2012, \mn@doi [\apj] {10.1088/0004-637X/750/1/10}, \href
  {https://ui.adsabs.harvard.edu/abs/2012ApJ...750...10S} {750, 10}

\bibitem[\protect\citeauthoryear{{Umetsu}, {Zitrin}, {Gruen}, {Merten},
  {Donahue}  \& {Postman}}{{Umetsu} et~al.}{2016}]{Umetsu2016}
{Umetsu} K.,  {Zitrin} A.,  {Gruen} D.,  {Merten} J.,  {Donahue} M.,
  {Postman} M.,  2016, \mn@doi [\apj] {10.3847/0004-637X/821/2/116}, \href
  {https://ui.adsabs.harvard.edu/abs/2016ApJ...821..116U} {821, 116}

\bibitem[\protect\citeauthoryear{{Varga} et~al.,}{{Varga}
  et~al.}{2019}]{varga2019}
{Varga} T.~N.,  et~al., 2019, \mn@doi [\mnras] {10.1093/mnras/stz2185}, \href
  {https://ui.adsabs.harvard.edu/abs/2019MNRAS.489.2511V} {489, 2511}

\bibitem[\protect\citeauthoryear{{Wright} \& {Brainerd}}{{Wright} \&
  {Brainerd}}{2000}]{wright2000}
{Wright} C.~O.,  {Brainerd} T.~G.,  2000, \mn@doi [\apj] {10.1086/308744},
  \href {http://adsabs.harvard.edu/abs/2000ApJ...534...34W} {534, 34}

\bibitem[\protect\citeauthoryear{{Wu}, {Weinberg}, {Salcedo}, {Wibking}  \&
  {Zu}}{{Wu} et~al.}{2019}]{wu2019}
{Wu} H.-Y.,  {Weinberg} D.~H.,  {Salcedo} A.~N.,  {Wibking} B.~D.,   {Zu} Y.,
  2019, \mn@doi [\mnras] {10.1093/mnras/stz2617}, \href
  {https://ui.adsabs.harvard.edu/abs/2019MNRAS.490.2606W} {490, 2606}

\bibitem[\protect\citeauthoryear{{Yang}, {Mo}, {van den Bosch}, {Jing},
  {Weinmann}  \& {Meneghetti}}{{Yang} et~al.}{2006}]{yang2006}
{Yang} X.,  {Mo} H.~J.,  {van den Bosch} F.~C.,  {Jing} Y.~P.,  {Weinmann}
  S.~M.,   {Meneghetti} M.,  2006, \mn@doi [\mnras]
  {10.1111/j.1365-2966.2006.11091.x}, \href
  {http://adsabs.harvard.edu/abs/2006MNRAS.373.1159Y} {373, 1159}

\bibitem[\protect\citeauthoryear{{Zhang} et~al.,}{{Zhang}
  et~al.}{2019}]{zhang2019}
{Zhang} Y.,  et~al., 2019, \mn@doi [\mnras] {10.1093/mnras/stz1361}, \href
  {https://ui.adsabs.harvard.edu/abs/2019MNRAS.487.2578Z} {487, 2578}

\bibitem[\protect\citeauthoryear{{Zitrin}, {Broadhurst}, {Bartelmann},
  {Rephaeli}, {Oguri}, {Ben{\'\i}tez}, {Hao}  \& {Umetsu}}{{Zitrin}
  et~al.}{2012}]{zitrin2012}
{Zitrin} A.,  {Broadhurst} T.,  {Bartelmann} M.,  {Rephaeli} Y.,  {Oguri} M.,
  {Ben{\'\i}tez} N.,  {Hao} J.,   {Umetsu} K.,  2012, \mn@doi [Monthly Notices
  of the Royal Astronomical Society] {10.1111/j.1365-2966.2012.21041.x}, \href
  {https://ui.adsabs.harvard.edu/abs/2012MNRAS.423.2308Z} {423, 2308}

\bibitem[\protect\citeauthoryear{{Zitrin} et~al.,}{{Zitrin}
  et~al.}{2013}]{zitrin2013}
{Zitrin} A.,  et~al., 2013, \mn@doi [The Astrophysical Journal]
  {10.1088/2041-8205/762/2/L30}, \href
  {https://ui.adsabs.harvard.edu/abs/2013ApJ...762L..30Z} {762, L30}

\makeatother
\end{thebibliography}

\appendix

\section{Covariance matrices}
\label{sec:cov}

\begin{figure}
    \begin{center}
    \includegraphics[width=\linewidth,angle=0.0]{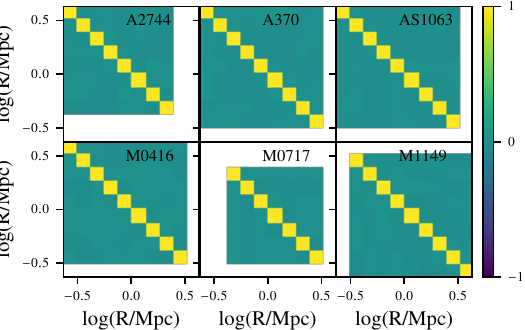}
    \caption{Correlation matrices for the six clusters. All off-diagonal terms are smaller than $3\%$.}
    \label{fig:corrmat}
    \end{center}
\end{figure}

As described in Sect.~\ref{sec:dsigma}, we assume in the analysis that covariances between the radial bins are negligible, and only include the variance when calculating the likelihood. Here we verify this assumption, and compute the full covariance matrices corresponding to the fiducial lensing signal, from the $100 \times 50$ resampled source catalogues described in Sect.~\ref{sec:dsigma}. The covariance between the $i^{\rm{th}}$ and $j^{\rm{th}}$ radial bin is then computed as:
\begin{equation}
    C_{ij} = \frac{1}{N}\sum_k^N \left(\Delta\Sigma_k(R_i) - \bar{\Delta\Sigma}(R_i)\right)^{\rm{T}}\left(\Delta\Sigma_k(R_j) - \bar{\Delta\Sigma}(R_j)\right),
\end{equation}
where $\Delta\Sigma_k$ is the lensing signal computed for the $k^{\rm{th}}$ source catalogue realisation, and $\bar{\Delta\Sigma}$ the mean over the $N = 5000$ realisations. For each cluster, off-diagonal terms are mcuh smaller than the diagonal ones. We quantify this by computing the correlation matrices, with $\rm{Corr}_{ij} = C_{ij}/\sqrt{Var_i Var_j}$, presented for each cluster in Fig.~\ref{fig:corrmat}. All off-diagonal terms are smaller than 3\%.

\label{lastpage}
\end{document}